\documentclass[10pt]{article}
\usepackage{amsmath}
\usepackage{graphicx}
\usepackage{amsfonts}
\usepackage{amssymb}
\usepackage{epsf}
\usepackage{latexsym}

\def\be{\begin{equation}}
\def\ee{\end{equation}}
\def\bea{\begin{eqnarray}}
\def\eea{\end{eqnarray}}
\def\Rho{\mbox{\Large $\rho$}}

\def\brho{\mbox{\boldmath$\rho$}}

\def\bPi{\mbox{\boldmath$\Pi$}}

\def\bGamma{\mbox{\boldmath$\Gamma$}}
\def\bsigma{\mbox{\boldmath$\sigma$}}
\def\btheta{\mbox{\boldmath$\theta$}}

\def\ttD{\mbox{\tt D}}

\def\ttH{\mbox{\tt H}}

\def\ttL{\mbox{\tt L}}

\def\ml{\mbox{\scriptsize l}}

\def\mn{\mbox{\scriptsize n}}

\def\NSI{Na\"{\i}ve Schr\"{o}dinger Interpretation }
\def\CPI{Conditional Probabilities Interpretation }
\def\NSII{Na\"{\i}ve Schr\"{o}dinger Interpretation}

\def\foo{\footnote}
\def\hat{\widehat}

\def\beq{\begin{equation}}
\def\eeq{\end{equation}}
\def\bea{\begin{eqnarray}}
\def\eea{\end{eqnarray}}
\def\pa{\partial}
\def\d{\textrm{d}}

\def\ttH{\mbox{\tt H}}
\def\ttL{\mbox{\tt L}}
\def\ttD{\mbox{\tt D}}

\def\5Star{\mbox{\Large$\star$}}

\def\cr{\mbox{\scriptsize{\bf $\mbox{ } \times \mbox{ }$}}}

\def\sumi3{\sum\mbox{}_{\mbox{}_{\mbox{\scriptsize $i$=1}}}^3}

\def\sumj3{\sum\mbox{}_{\mbox{}_{\mbox{\scriptsize $j$=1}}}^3}
\def\sumk3{\sum\mbox{}_{\mbox{}_{\mbox{\scriptsize $k$=1}}}^3}

\def\sss{\mbox{\scriptsize s}}

\def\ma{\mbox{a}}

\def\md{\mbox{d}} 
\def\me{\mbox{e}}

\def\mh{\mbox{h}}
\def\mi{\mbox{i}}

\def\ml{\mbox{l}}

\def\mn{\mbox{n}}

\def\mp{\mbox{p}}

\def\ms{\mbox{s}}

\def\muu{\mbox{u}}
 
\def\mw{\mbox{w}}

\def\mD{\mbox{D}}
\def\mE{\mbox{E}}

\def\mI{\mbox{I}}

\def\mK{\mbox{K}}
\def\mL{\mbox{L}}

\def\mO{\mbox{O}}
\def\mP{\mbox{P}}

\def\mR{\mbox{R}}
\def\mS{\mbox{S}}

\def\mZ{\mbox{Z}} 
\def\sa{\mbox{\scriptsize a}}

\def\sc{\mbox{\scriptsize c}}
\def\sd{\mbox{\scriptsize d}}
\def\se{\mbox{\scriptsize e}}
\def\sf{\mbox{\scriptsize f}}
 
\def\sh{\mbox{\scriptsize h}} 
\def\si{\mbox{\scriptsize i}}

\def\sll{\mbox{\scriptsize l}}  
\def\sm{\mbox{\scriptsize m}}
\def\sn{\mbox{\scriptsize n}} 
 
\def\sp{\mbox{\scriptsize p}}
\def\sq{\mbox{\scriptsize q}}
\def\sr{\mbox{\scriptsize r}}

\def\sB{\mbox{\scriptsize B}}
\def\sC{\mbox{\scriptsize C}}
\def\sD{\mbox{\scriptsize D}}

\def\sI{\mbox{\scriptsize I}}
\def\sJ{\mbox{\scriptsize J}}
\def\sK{\mbox{\scriptsize K}}

\def\sO{\mbox{\scriptsize O}}

\def\sR{\mbox{\scriptsize R}}
\def\sS{\mbox{\scriptsize S}}

\def\sW{\mbox{\scriptsize W}}

\def\barp{\bar{\tt p}}
\def\barq{\bar{\tt q}}
\def\barr{\bar{\tt r}}
\def\ttH{\mbox{\tt{H}}}

\def\sC{\mbox{\scriptsize C}}
\def\eph{\mbox{\scriptsize eph}}
 
\def\eph(B){\mbox{\scriptsize em(JBB)}}

\def\eph(B){\mbox{\scriptsize emergent(JBB)}}

\def\tK{\mbox{\tiny K}}

\def\tR{\mbox{\tiny R}}
\def\tS{\mbox{\tiny S}}

\def\bfM{\mbox{{\bf \sffamily M}}}

\def\bK{\mbox{{\bf K}}}

\def\bR{\mbox{{\bf R}}}

\def\bn{\mbox{{\bf n}}}
\def\bq{\mbox{{\bf q}}}
\def\br{\mbox{{\bf r}}}

\def\by{\mbox{{\bf y}}}

\def\fG{\mbox{\sffamily G}}
\def\fH{\mbox{\sffamily H}}

\def\fQ{\mbox{\Large $\mathfrak{q}$}}

\def\sFG{\mbox{$\mathfrak{g}$}}

\def\fG{\mbox{\Large $\mathfrak{g}$}}

\def\sfa{\mbox{\sffamily{\scriptsize a}}}
\def\sfb{\mbox{\sffamily{\scriptsize b}}}

\def\sfg{\mbox{\sffamily{\scriptsize g}}}

\def\sfA{\mbox{\sffamily{\scriptsize A}}}
\def\sfB{\mbox{\sffamily{\scriptsize B}}}

\def\sfG{\mbox{\sffamily{\scriptsize G}}}

\def\sfZ{\mbox{\sffamily{\scriptsize Z}}}
\newcommand{\Prob}{{\rm Prob}}

\def\K{Kucha\v{r} }

\def\bn{\mbox{\bf n}}
\def\bp{\mbox{\bf p}}

\def\bP{\mbox{\bf P}}

\def\bQ{\mbox{\bf Q}}
\def\bS{\mbox{\bf S}}

\begin{document}
\begin{titlepage}

\vspace{.7in}

\begin{center}

\LARGE{\bf APPROACHING THE PROBLEM OF TIME}\normalsize  

\vspace{.1in}

\large{\bf WITH A COMBINED SEMICLASSICAL-RECORDS-HISTORIES SCHEME}\normalsize 

\normalsize

\vspace{.4in}

{\large \bf Edward Anderson$^*$}

\vspace{.2in}

\large {\em APC AstroParticule et Cosmologie, Universit\'{e} Paris Diderot CNRS/IN2P3, CEA/Irfu, 
Observatoire de Paris, Sorbonne Paris Cit\'{e}, 10 rue Alice Domon et L\'{e}onie Duquet, 75205 Paris Cedex 13, France.} 

\vspace{.15in} 

and {\em DAMTP, Centre for Mathematical Sciences, Wilberforce Road, Cambridge CB3 OWA} \normalsize

\begin{abstract}

I approach the Problem of Time and other foundations of Quantum Cosmology using a combined histories, timeless and semiclassical approach.
This approach is along the lines pursued by Halliwell.  
It involves the timeless probabilities for dynamical trajectories entering regions of configuration space, which are computed within 
the semiclassical regime.
Moreover, the objects that Halliwell uses in this approach commute with the Hamiltonian constraint, H.  
This approach has not hitherto been considered for models that also possess nontrivial linear constraints, Lin. 
This paper carries this out for some concrete relational particle models (RPM's).
If there is also commutation with Lin -- the \K observables condition -- the constructed objects are Dirac observables.  
Moreover, this paper shows that the problem of \K observables is explicitly resolved for 1- and 2-$d$ RPM's.   
Then as a first route to Halliwell's approach for nontrivial linear constraints that is also a construction of Dirac observables, 
I consider theories for which \K observables are formally known, giving the relational triangle as an example.  
As a second route, I apply an indirect method that generalizes both group-averaging and Barbour's best matching.    
For conceptual clarity, my study involves the simpler case of Halliwell 2003 sharp-edged window function.   
I leave the elsewise-improved softened case of Halliwell 2009 for a subsequent Paper II.  
Finally, I provide comments on Halliwell's approach and how well it fares as regards the various facets of the 
Problem of Time and as an implementation of QM propositions. 

\end{abstract}

\end{center}
     
\vspace{1in}   

\vspace{0.2in}

\vspace{3in}

\noindent $^*$ eanderso@apc.univ-paris7.fr 

\end{titlepage}

\section{Introduction}

\subsection{The Problem of Time and its Frozen Formalism Facet} \label{SSec: 1.6}

This notorious problem \cite{DeWitt67, Battelle, Kuchar81, Kuchar91, Kuchar92, I93, Kuchar93, Kuchar99, Rovellibook, Kieferbook, Smolin08, 
APOT, FileR, APOT2} occurs because the `time' of GR and the `time' of ordinary Quantum Theory are mutually incompatible notions.
This incompatibility leads to a number of problems with trying to replace these two branches with a 
single framework in situations in which the premises of both apply, such as in black holes and in the very early universe.  
This is investigated in this paper via relational particle models (RPM's) \cite{BB82, B94I, B94IIEOT, B03, 06I, 06II, SemiclI, TriCl, FORD, Cones,  
+Refs, Records12, 08I, 08II, AF, +Tri, ScaleQM, 08III, SemiclIII, QShapeQSub, QuadI, QuadII, FileR, ARel, ARel2, ACos2}.  
Sec 1.5 explains the nature of these models. 
For now, I note that they have a large number of analogies with GR (see \cite{Kuchar92, B94I, B94IIEOT, Kieferbook, 06II, Smolin08, 08II, AF, 
+Tri, ScaleQM, 08III, SemiclIII} and especially \cite{FileR}), rendering them suitable as toy models for many aspects of the Problem of Time.

One notable facet of the Problem of Time shows up in attempting canonical quantization of GR (or of many other gravitational theories that carry 
over likewise background-independent).
It is due to GR's {\it Hamiltonian constraint}\foo{The spatial 
topology M is taken to be compact without boundary. 
$h_{\mu\nu}$ is a spatial 3-metric thereupon, with determinant $h$, covariant derivative $D_{\mu}$, 
Ricci scalar Ric($h$) and conjugate momentum $\pi^{\mu\nu}$.  
$\Lambda$ is the cosmological constant.
Here, the GR configuration space metric ${\cal M}^{\mu\nu\rho\sigma} = h^{\mu\rho}h^{\nu\sigma}$ 

\noindent  $- h^{\mu\nu}h^{\rho\sigma}$ is the undensitized inverse DeWitt supermetric with determinant ${\cal M}$ and inverse ${\cal N}_{\mu\nu\rho\sigma}$ 
itself the undensitized DeWitt supermetric $h_{\mu\rho}h_{\nu\sigma} - h_{\mu\nu}h_{\rho\sigma}/2$. 
In this paper, $\lfloor \mbox{ } \mbox{ } \rfloor$ is a portmanteau of function dependence ( \mbox{ } ) in finite theories such as 
particle mechanics or minisuperspace and functional 
dependence [ \mbox{ } ] or mixed function-functional dependence ( \mbox{ } ; \mbox{ } ] in infinite theories such as field theories. 
I use $\int_{\tS}\md \mS$ for the integral over space, including the finite case, for which this is taken to be a multiplicative 1.}  
\beq
{\cal H} := {\cal N}_{\mu\nu\rho\sigma}\pi^{\mu\nu}\pi^{\rho\sigma}/\sqrt{h} - \sqrt{h}\{\mbox{Ric}(x;h] - 2\Lambda\}  = 0  \mbox{ }   
\eeq
being quadratic but not linear in the momenta; I denote the general case of such a constraint by ${\cal Q}$uad.

Then promoting a constraint with a momentum dependence of this kind to the quantum level gives a time-independent wave equation 
\beq
\hat{\cal H}\Psi = 0
\label{WDE} \mbox{ } ,   
\eeq
in place of ordinary QM's time-dependent one, 
\beq
i\pa\Psi/\pa t = \hat{\fH}\Psi \mbox{ } 
\eeq 
[or a functional derivative $\delta/\delta t(x^{\mu})$ counterpart of this in the general GR case].  
Here, $\fH$ is a Hamiltonian, $\Psi$ is the wavefunction of the universe and $t$ is absolute Newtonian time [or a a local GR-type generalization
$t(x^{\mu})$].

In the case of GR, in more detail, (\ref{WDE}) is a Wheeler--DeWitt equation,\footnote{The inverted commas indicate that the Wheeler-DeWitt equation 
has, in addition to the Problem of Time, various technical problems, including 

\noindent A)  regularization problems -- not at all straightforward for an equation for a theory of an infinite number of degrees of freedom in the absense of background structure, while the mathematical meaningfulness of functional differential equations is open to question.
N.B. this is not an issue in the specific examples in this paper as these are for a finite number of degrees of freedom. 

\noindent B) There are operator-ordering issues, which this paper's toy models do exhibit an analogue of \cite{Banal, FileR}. 
These are tied to the number $\xi$, and in the relational approach I resolve this \cite{Banal} by taking the conformal operator-ordering option \cite{DeWitt57, Magic, ConfOrder, HP86}, 
which is underlied by the relational formulation of the action (see Sec 1.5).}
\beq
\hat{\cal H}\Psi := - \mbox{`}\left\{\frac{1}{\sqrt{{\cal M}}}\frac{\delta}{\delta h^{{\mu\nu}}}
\left\{
\sqrt{{\cal M}}{\cal N}^{\mu\nu\rho\sigma}\frac{\delta\Psi}{\delta h^{{\rho\sigma}}}
\right\} 
- \xi \,\mbox{Ric}(h;{\cal M}]\right\}\mbox{'}\,\Psi -  \sqrt{h}\mbox{Ric}(x;h]\Psi + {\sqrt{h}2\Lambda   }\Psi  = 0  \label{WDE3} \mbox{ }  .      
\eeq
This suggests, in apparent contradiction with everyday experience, that nothing at all happens in the universe! 
Thus one is faced with having to explain the origin of the notions of time in the laws of physics that appear to apply in the universe.  
Sec 1.3 considers a number of strategies toward explaining this.   
[Moreover timeless equations such as the Wheeler--DeWitt equation apply {\sl to the universe as a whole}, whereas the more ordinary laws of physics apply to 
small subsystems {\sl within} the universe, suggesting this to be an apparent `paradox' rather than an actual one.]

\subsection{Other Facets of the Problem of Time} 

\noindent Over the past decade, it has become more common to suggest or imply that the Problem of Time {\sl is} the Frozen Formalism Problem.  
However, a more long-standing point of view \cite{Kuchar91, Kuchar92, I93, Kuchar93} (and also argued in favour for in e.g. \cite{Kuchar99, Kieferbook, APOT, FileR, APOT2} 
and the present article) is that the Problem of Time contains a number of further facets.  
Furthermore, this is not even a case of then having to address around eight facets in succession.
For, as Kucha\v{r} found, these facets interfere with each other.  
He termed this \cite{Kuchar93} a `many gates' problem, in which one attempting to enter the gates in sequence finds that they are no 
longer inside some of the gates they had previously entered. 
I argue that this occurs because the facets arise from a common cause -- the mismatch of the notions of time in GR and Quantum Theory -- 
by which they bear conceptual and technical relations, making it advantageous, and likely necessary for genuine progress, to treat them 
as a coherent package rather than piecemeal as `unrelated problems'.  
To emphasize this common cause, and place the eight principal facets into a more memorable form (I have found most theoreticians to be 
unaware of some facets and/or of their time connotations), I used an even more vivid mnemonic \cite{APOT, FileR}, depicting the eight 
facets as the breath, wings, tail, scales and four sets of claws that constitute a single entity: the Ice Dragon.  
This depiction makes it clear that the eight facets inter-protect each other (as Kucha\v{r} observed), so that resolving the Problem of 
Time is altogether harder than addressing each of the facets in succession, let alone just unfreezing the equations of physics.

Apart from the `frozen breath', only the `wings' of the Problem of Observables shall play a recurring theme in the present article. 
\noindent The Frozen Formalism Facet itself has the following addendum.  

\mbox{ } 

\noindent {\bf Hilbert Space/Inner Product Problem}, i.e. how to turn the space of solutions of the frozen equation in question into a Hilbert space.
In GR, indefiniteness of the kinetic metric prevents use of a Schr\"{o}dinger inner product and then a Klein--Gordon type one 
fails on other grounds \cite{Kuchar81}.  
This is a time problem since inner products and unitary evolution are tied.  

\mbox{ } 
 
\noindent We next require an interlude so as to interpret further facets.  
GR also has a linear {\it momentum constraint} 
\beq
{\cal H}_{\mu} := -2D_{\nu}\pi^{\nu}\mbox{}_{\mu} = 0 \mbox{ } \mbox{ which becomes, at the quantum level, } \mbox{ } 
\widehat{\cal H}_{\mu}\Psi := -2D_{\rho}h^{\rho\nu}\delta\Psi/\delta h_{\mu\nu} = 0 \mbox{ } ;
\label{5}
\eeq 
this often becomes entwined in technical problems that afflict Problem of Time strategies.  

\noindent 
Many such entwinings (see e.g. the next two Problems below) furthermore specifically concern diffeomorphisms [whether  i) the 
spatial diffeomorphisms associated with the GR momentum constraint itself, ii) the spacetime diffeomorphisms or iii) the result 
splitting the spacetime diffeomorphisms due to a foliation by spatial hypersurfaces as occurs in canonical approaches to GR].  
%
%
iii) is altogether harder to handle; whilst i) and ii) are both infinite-dimensional Lie groups, in iii) the GR Hamiltonian constraint 
does not directly account for the diffeomorphisms of ii) that are not present amongst those of i) 
This is reflected e.g. in the Dirac algebra of the GR constraints having not the mere structure constants of a Lie algebra but, 
rather, structure functions.
%
%
At least this algebra does retain the useful property of closing at the classical level without further constraints appearing.
\noindent 
Additionally, in classical GR, one can foliate spacetime in many ways, each corresponding to a different choice of timefunction.  
This is how time in classical GR comes to be `many-fingered', with each finger `pointing orthogonally' to each foliation.  
Finally, classical GR has the remarkable property of being foliation-independent \cite{HKT}, so that going between two given spatial 
geometries by means of different foliations in between produces the same region of spacetime, thus giving the same answers to whatever 
physical questions can be posed therein.  
Equipped with this knowledge, one can return to the discussion of the Problem of Time.

\mbox{ }  

\noindent {\bf Best Matching Problem}: I argued in \cite{FileR, ARel, APOT2} that this is a more general conceptualization of, 
and name for, the usually-listed {\bf Sandwich Problem}. 
The context for this is that, if the theory has $\fG$-nontriviality for $\fG$ the group associated with the linear constraints 
${\cal L}\mi\mn_{\sfZ}$ [generalization of (\ref{5})], then varying with respect to the auxiliary $\fG$-variables produces ${\cal L}$in$_{\sfZ}$. 
The Best Matching Problem is then whether one can solve the Lagrangian form of ${\cal L}\mi\mn_{\sfZ} = 0$ for the $\fG$-auxiliaries themselves, 
which is a particular form of reduction.
This problem is clearly already present at the classical level.
It furthermore becomes entwined in some of the approaches that start with manipulations at the classical level prior to quantizing.

\mbox{ } 

\noindent {\bf Foliation Dependence Problem}: for all that one would like a quantization of GR to retain the nice classical property of 
refoliation invariance, at the quantum level there ceases to be an established way of guaranteeing this property.    
That this is obviously a time problem follows from each foliation by spacelike hypersurfaces being orthogonal to a GR timefunction: 
each slice corresponds to an instant of time for a cloud of observers distributed over the slice.
Each foliation corresponds to such a cloud of observers moving in a particular way.   

\mbox{ }

\noindent {\bf Functional Evolution Problem} alias {\bf Possibility of Anomalies}. 
This is the quantum-level problem that the commutator of constraints is capable of manifesting, involving breakdown of the 
immediate closure of the constraint algebra that occurs at the classical level.  
Dirac \cite{Dirac} said that one had to be lucky so as to avoid such breakdowns.  
Note that only some of the anomalies that one finds in physics are time-related.
However, in the present case of the quantum counterpart of the Dirac algebra of GR constraints, this is a time 
issue due to what these constraints signify.
In particular, non-closure here is a way in which the Foliation Dependence Problem can manifest itself, 
through the non-closure becoming entwined with details of the foliation. 
[Non-closure is also entwined with the Operator Ordering Problem, since changing the operator ordering gives 
additional right-hand-side pieces not present in the classical Poisson brackets algebra.]  

\mbox{ } 

\noindent {\bf Multiple Choice Problem} alias {\bf Kucha\v{r}'s Embarrassment of Riches} \cite{Kuchar92}.  
This is the purely quantum-mechanical problem that different choices of time variable may give inequivalent quantum theories.  
[The `riches' are then the multiplicity of such inequivalent quantum theories.]  
\noindent Foliation Dependence is one of the ways in which the Multiple Choice Problem can manifest itself.
Moreover, the Multiple Choice Problem is known to occur even in some finite toy models, so that 
foliation issues are not the only source of the Multiple Choice Problem.  
For instance, another way the Multiple Choice Problem can arise is as a subcase of how making different 
choices of sets of variables to quantize may give inequivalent quantum theories, as follows from e.g. the Groenewold--van Hove phenomenon.  

\mbox{ } 

\noindent {\bf Global Problem of Time} alias {\bf Kucha\v{r}'s Embarrassment of Poverty} \cite{Kuchar92}.  
This is already present at the classical level. 
The spatial part of this problem consists in the separation into true variables and embedding (space frame and timefunction) variables' having 
a proneness for being globally impossible, for reasons that closely parallel the Gribov effect in Yang--Mills theory.  
This proneness to not globally exist explains Kucha\v{r}'s name for this problem. 
The temporal part of this problem concerns cases in which such a split can be defined but cannot be indefinitely 
continued as one progresses along a foliation (or simpler models' timefunctions eventually going astray).  


\noindent {\bf Spacetime (Reconstruction or Replacement) Problem} \cite{Kuchar92, APOT2}.    
Internal space or time coordinates to be used in the conventional classical spacetime context need to be 
scalar field functions on the spacetime 4-manifold.    
In particular, these do not have any foliation dependence.
However, the canonical approach to GR uses functionals of the canonical variables, and which there is no a priori reason for such to be scalar 
fields of this type.  
Thus one is faced with either finding functionals with this property, or coming up with a new means of 
reducing to the standard spacetime meaning at the classical level.   
There are further issues involving properties of spacetime being problematical at the quantum level.  
Quantum Theory implies fluctuations are unavoidable, but now that these amount to fluctuations of 
3-geometry, they are too numerous to fit within a single spacetime (see e.g. \cite{Battelle}).  
Thus (something like) the superspace picture (considering the set of possible 3-geometries) 
might take over from the spacetime picture at the quantum level.  
It is then not clear what becomes of causality (or of locality, if one believes that the quantum 
replacement for spacetime is `foamy' \cite{Battelle}).  
There is also an issue of recovering continuity in suitable limits in approaches that treat space 
or spacetime as discrete at the most fundamental level (see e.g. \cite{Disc}).  

\mbox{ }

\noindent  {\bf Problem of Beables}. This is usually called the Problem of Observables \cite{Dirac, Kuchar92, I93, Rovellibook}, but I follow 
Bell \cite{Speakable, Bell} in ascribing more physical significance to beables than to observables, especially in the context of Quantum Cosmology.
I prefer to use beables, or yet other names corresponding to local versions of such concepts, as per Sec 8.2.
The Problem of Beables involves construction of a sufficient set of beables for the physics of one's model; there are then 
involved in the model's notion of evolution.  
The Problem of Beables was depicted \cite{FileR} as the `wings' of the Ice Dragon, due to observables/beables issues popping up all over the 
place in Theoretical Physics, much like that the Ice Dragon suddenly popping up in unexpected places of the realm through having wings to 
swiftly and unexpectedly fly there).  
Though, as we shall see in Sec 2, `whether the Ice Dragon has wings' remains a debated point.

\subsection{A selection of Problem of Time strategies}

\noindent A) {\bf Semiclassical Approach}.  
Perhaps one has slow, heavy `h'  variables that provide an approximate timestandard with respect to which the other fast, light `l' 
degrees of freedom evolve \cite{HallHaw, Kuchar92, Kieferbook}.  
In the Halliwell--Hawking \cite{HallHaw} scheme for GR Quantum Cosmology, $h$ is scale (and homogeneous matter modes) and the l-part 
are small inhomogeneities.\footnote{This is a quantum cosmological explanation for the origin of structure in the universe --  
the seeding of galaxies and of CMB inhomogeneities).  
The connection between quantum-cosmological perturbations and the observed universe is usually via some inflationary mechanism.}  
I.e., it is a perturbative midisuperspace model. 
The Semiclassical Approach involves making
\beq
\mbox{i) the Born--Oppenheimer (BO) ansatz} \hspace{1in}  \Psi(\mh, \ml) = \psi(\mh)|\chi(\mh, \ml)\rangle \mbox{ } , \hspace{4in}  
\label{BO}
\eeq  
\beq
\mbox{ii) the WKB ansatz} \hspace{2.15in}
\psi(\mh) = \mbox{exp}(i\, S(\mh)) \mbox{ } ; \hspace{5in}  
\label{WKB}
\eeq 
in each case one makes a number of associated approximations.    
\noindent [Evoking semiclassicality only makes sense if one's goals are somewhat modest as compared to some more general goals 
in Quantum Gravity programs.  
Nevertheless this is still useful for some applications, including the foundations of practical Quantum Cosmology i.e. putting calculations 
along the lines of Halliwell and Hawking's.]   


\noindent iii) One forms the h-equation,  
$
\langle\chi| \hat{\cal H} \Psi = 0.  
$
\noindent Then, under a number of simplifications, this yields a Hamilton--Jacobi equation\foo{For simplicity, I 
present this in the case of 1 h degree of freedom and with no linear constraints.} 
$\{\pa S/\pa\mh\}^2 = 2\{E - V(\mh)\}$ , where $V(\mh)$ is the h-part of the potential and $E$ is the total energy.

One way of solving this involves doing so for an approximate emergent semiclassical time $t^{\se\sm(\sW\sK\sB)}(\mh)$; 
this happens to be \cite{SemiclI, FileR} a recovery of a `time before quantization' timefunction: the Jacobi--Barbour--Bertotti one 
\cite{BB82, B94I}.    


\noindent iv) One then forms the l-equation 
$\{1 - |\chi\rangle\langle\chi|\}\hat{\cal H}\Psi = 0$. 
This is a fluctuation equation but it can be recast (modulo further approximations) into a 
$t^{\se\sm(\sJ\sB\sB)}$-dependent Schr\"{o}dinger equation for the l-degrees of freedom, 
\beq
i\pa|\chi\rangle/\pa t^{\se\sm}  = \widehat{\fH}_{\sll}|\chi\rangle \mbox{ }
\label{TDSE2} \mbox{ } .    
\eeq
The emergent time dependent left-hand side of this arises from the cross-term $\pa_{\sh}|\chi\rangle\pa_{\sh} \psi$. 
$\widehat{\fH}_{\sll}$ is the remaining surviving piece of $\widehat{\cal H}$, acting as a Hamiltonian for the l-subsystem.  
Moreover, the working leading to such a time-dependent wave equation ceases to function in the absence of making the WKB ansatz and approximation, 
which, additionally, in the quantum-cosmological context, is not known to be a particularly strongly supported ansatz and approximation to make.    

\mbox{ } 

\noindent B) {\bf Timelessness}. A number of approaches take timelessness at face value.  
One considers only questions of the universe `being', rather than `becoming', a certain state.  


\noindent B.1) A first example is the {\it \NSII}. 
(This is due to Hawking and Page \cite{HP86}, though its name was coined by Unruh and Wald \cite{UW89}.)  
This concerns the `being' probabilities for universe properties such as: what is the probability that the universe is large? 
Flat? 
Isotropic? 
Homogeneous?   
One proceeds via considering the probability that the universe belongs to region R of the 
configuration space that corresponds to a quantification of a particular such property, 
\beq
\mbox{Prob}(\mR) \propto \int_{\sR}|\Psi|^2\mathbb{D}\bQ \mbox{ } , 
\eeq 
for $\mathbb{D} \bQ$ the configuration space volume element.
This approach is termed `na\"{\i}ve' due to it not using any further features of the constraint equations.  
Its implementation of propositions is Boolean via how the region enters the integral.
%


\noindent B.2) The {\it \CPI} \cite{PW83} goes further by addressing conditioned questions of `being'. 
The conditional probability of finding $B$ in the range $b$, given that $A$ lies in $a$, and to allot to it the value 
\beq
\Prob(B\in b | A\in a; \Rho) = \frac{\mbox{tr}
\big(
\mP^B_{b}\,\mP^A_{a}\,\Rho\,\mP^A_a
\big)                            }{
\mbox{tr}
\big(
\mP^A_a\,\Rho
\big)
} \mbox{ } ,    
\label{CPI}
\eeq
where $\Rho$ is a density matrix for the state of the system and the $\mP^A_a$ denote projectors.  
An example of such a question is `what is the probability that the universe is flat given that it is isotropic'?  
This approach can additionally deal with the question of `being at a time' by having the conditioning proposition refer to a `clock' subsystem.  


\noindent B.3) {\it Records Theory}  involves localized subconfigurations of a single instant of time.  
It concerns issues such as whether these contain useable information, are correlated to each other, and 
whether a semblance of dynamics or history arises from this (which can involve reducing questions of `becoming' to questions of `being').
Records Theory is a heterogeneous subject, a number of different authors advocating different notions of Records Theory 
are Page \cite{Page12}, Barbour \cite{B94IIEOT} and I \cite{Records12, FileR, ARel2} (see also the next SSec).

\mbox{ }

\noindent C) {\bf Histories Theory}. 
Perhaps instead it is the histories that are primary, a view brought to the GR 
context by Gell--Mann and, especially, Hartle  \cite{GMH, Hartles} and subsequently worked on by Isham, Linden, Savvidou and others \cite{IL2, IL, ILConcat}.

\subsection{Halliwell's combined approach}

Some motivation for this is as follows.   


\noindent Pro 1) Both histories and timeless approaches lie on the common ground of atemporal logic structures \cite{IL, FileR}.

\noindent Pro 2) There is a Records Theory within Histories Theory, as pointed out by Gell-Mann and Hartle \cite{GMH} and 
further worked on by them \cite{GMH11}, and by Halliwell and collaborators \cite{H99, HT, H03, HT01HDHW, ATOverlap, H09, H11}.
Thus Histories Theory supports Records Theory by providing guidance as to the form a working Records Theory would take.
This also allows for these two to be jointly cast as a mathematically-coherent package (as already illustrated in preceding subsections).
As Gell-Mann and Hartle say \cite{GMH},
\beq
\mbox{Records are ``{\it somewhere in the universe where information is stored when histories decohere}"}. 
\label{situ}
\eeq
Anti 1) There is some chance that being more minimalistic than Gell-Mann--Hartle and Halliwell could itself succeed.  
If one alternatively considers records from first principles from which histories are to be derived, as the history collapses to a single instant, 
path integrals cease to be defined so this kind of decoherence functional is absent from Records Theory.

\noindent Pro 3) Histories decohereing is a leading (but as yet {\sl not fully established}) way by which a semiclassical regime's 
WKB approximation could be legitimately obtained in the first place.  
Thus Histories Theory could support the Semiclassical Approach by freeing it of a major weakness.

\noindent Pro 4) The Semiclassical Approach and/or Histories Theory could plausibly support Records Theory by providing a mechanism 
for the semblance of dynamics \cite{H03, Kieferbook} (though the possibility of a practically useable such occurring within pure 
Records Theory has not been overruled).  
Such would go a long way towards Records Theory being complete.  
N.B. that emergent semiclassical time amounts to an approximate semiclassical recovery \cite{SemiclI} of Barbour's classical emergent time \cite{B94I, ARel2, ACos2}, 
which is an encouraging result as regards making such a Semiclassical--Timeless Records combination.

\noindent Pro 5) The elusive question of which degrees of freedom decohere which should be answerable through where in 
the universe the information is actually stored, i.e. where the records thus formed are \cite{GMH, H03}.  
In this way, Records Theory could in turn support Histories Theory.

\noindent Pro 6) The Semiclassical approach aids in the computation of timeless probabilities of histories entering given configuration 
space regions.
This is by the WKB assumption giving a semiclassical flux into each region \cite{H03} in terms of $S(\mh)$ and the Wigner function (defined in 
Sec 3).  
Such schemes go beyond the standard Semiclassical Approach, and as such there may be some chance that further objections to the Semiclassical Approach 
(problems inherited from the Wheeler--DeWitt equation and Spacetime Reconstruction Problems) would be absent from the new unified strategy.  

\noindent Pro 7) Halliwell's approach has further foundational value for practical Quantum Cosmology.  


\noindent For details of the parts of Halliwell's program considered in this paper, see Sec 3.    

\mbox{ } 

\noindent Note 1) This program does not make further use of the Semiclassical Approach's  specific unfreezing moves, 
for all that it does use BO-WKB approximations in obtaining various expressions relevant to the Problem of Time.

\noindent Note 2) Halliwell's approach's attitude to observables/beables is a Dirac one (see Sec 2), though he does not for now consider the 
effect of linear constraints;  the present paper's study is the first to consider examples possessing linear constraints.

\subsection{Outline of relational particle mechanics (RPM's)  }\label{RPMOut}

\noindent RPM's are whole-universe models, toy models of geometrodynamics, with nontrivial constraints and structure formation.   
\noindent These are suitable for the study of Quantum Cosmology and some Problem of Time aspects.
\noindent In particular, they are a good arena for Histories, Records and Semiclassical Approaches, and quite possibly observables/beables based 
approaches too (a point toward which the present paper contributes).
\noindent 
RPM's obey the following `Leibniz--Mach--Barbour' brand of relationalism (GR can be cast in a closely parallel form too \cite{RWR, FileR}).

\mbox{ } 

\noindent A physical theory is {\bf temporally relational} if there is {\sl no meaningful primary notion of time for the system as a whole} 
(e.g. the universe) \cite{BB82, RWR, FORD, ARel}.  
\noindent The very cleanest implementation of this is by using actions that are 
\noindent i) {\sl manifestly parametrization irrelevant}, and
\noindent ii) {\sl free of extraneous time-related variables}. 
[E.g. one is not to involve external absolute Newtonian time, or the dynamical formulation of GR's `lapse' variable.]   
The relational action that implements the above in the case of RPM's is the Jacobi-type action
\beq
S = \sqrt{2} \int \d s\sqrt{E - V} \mbox{ } , 
\label{Actio}
\eeq
for $\d s$ the kinetic arc element. 

\mbox{ } 

\noindent A physical theory is {\bf configurationally relational} with respect to a group of transformations $\fG$ if configurations for the 
whole universe inter-related by a $\fG$-transformations are indiscernible \cite{BB82, B03, RWR, LanPhan, FORD, FEPI, Cones, ARel}.    
One way to implement this is to use not `bare' configurations but rather their arbitrary-$\fG$-frame-corrected counterparts.
This is the only known way, with sufficiently widespread applicability for a relational program to underlie the 
whole of the classical fundamental physics `status quo' of GR coupled to the Standard Model \cite{RWR}.
Moreover, this paper also covers RPM cases for which configurational relationalism is directly implementable.

\mbox{ } 

The following constraints arise from these relational postulates.  
The energy-type constraint
\beq
{\cal E} := \sum\mbox{}_{\mbox{}_{\mbox{\scriptsize $I$ = 1}}}^{N} p_I^2/2m_I +  V(q^I) = E \mbox{ }  
\label{EEE}
\eeq
results from the parametrization-irrelevance of the action.  
Note that this is purely quadratic in the momenta.
On the other hand, the zero total angular momentum constraint 
\beq
\mbox{\boldmath${\cal L}$} := \sum\mbox{}_{\mbox{}_{\mbox{\scriptsize $I$ = 1}}}^{N} \bq^I \cr \bp_I = 0  
\label{ZAM}
\eeq
results from variation with respect to the auxiliaries that indirectly impose the rotational part of the configurational relationalism.
In the case of pure-shape RPM, this is accompanied by the zero total dilational constraint
\beq
{\cal D} := \sum\mbox{}_{\mbox{}_{\mbox{\scriptsize $I$ = 1}}}^{N} \bq^I \cdot \bp_I = 0 \mbox{ }   
\eeq
that results from variation with respect to the auxiliaries that indirectly impose the dilational part of the configurational relationalism. 
Clearly these last two constraints are linear in the momenta.

\noindent By the purely quadratic form of ${\cal E}$  the frozen formalism facet of the Problem of Time clearly manifests itself for RPM's.  

The above is an indirect, eventually Dirac, presentation, but there is also an explicit r-presentation (the coincidence of reduced representation 
\cite{+Tri, FileR} -- linear constraints eliminated -- and of mechanics set up directly on the reduced configuration space \cite{FORD, Cones, FileR} 
-- linear constraints never formulated) in dimensions 1 and 2. 
Here then the only constraint is 
\beq
{\cal E}_{\sr} := N^{\sfA\sfB}P_{\sfA}P_{\sfB}/2 + V(Q^{\sC}) = E \mbox{ }  
\eeq
for configuration space coordinates $Q^{\sfA}$ with conjugate momenta $P_{\sfA}$ and configuration space metric $M_{\sfA\sfB}$ with inverse $N^{\sfA\sfB}$.  
\noindent \cite{FileR} gives 106 similarities and 39 differences between RPM's and geometrodynamics as motivation.    
%
%
\noindent RPM's particular midisuperspace features are a notion of clumping/inhomogeneity/structure and possessing linear constraints; 
both of these features are of especial relevance to the foundations of Quantum Cosmology, rendering RPM's as good qualitative toy model for the Halliwell--Hawking approach.

\subsection{The status of the Problem of Time facets for RPM's}

\noindent The {\sl Frozen Formalism Facet for RPM's} is present via the purely quadratic dependence of (\ref{EEE}) on the momenta.  

\noindent There is {\sl (mostly) no inner product problem for RPM's}: positive-definite mechanics, so the Schr\"{o}dinger inner product suffices.


\noindent  The {\sl Best Matching Problem for RPM's} is  for 1- and 2-$d$ RPM's, not a problem but rather a {\sl resolved} situation 
\cite{FORD, Cones, FileR} that opens up extra  paths and checks not available for geometrodynamics.   


\noindent There is {\sl no Foliation Dependence Problem for RPM's}, since the foliation 
of spacetime/embedding into spacetime meaning of GR's Dirac Algebra of constraints is lost through toy-modelling 
it with rotations (and/or dilations).  


\noindent There is {\sl no Functional Evolution Problem for RPM's}: these are `lucky' in Dirac's sense \cite{FileR}.     


\noindent The {\sl Multiple Choice Problem for RPM's}.  
This is already present for finite systems \cite{Kuchar92, I93}, and occurs for RPM's much as it does in minisuperspace \cite{I93}.    


\noindent The {\sl Global Problem for RPM's}.  This is present (see \cite{FileR} for details), including, unlike for minisuperspace, RPM's having 
spatial globality issues via RPM's possessing meaningful notions of localization/clumping.  


\noindent There is {\sl no Spacetime Reconstruction Problem for RPM's} since they have no nontrivial notion of spacetime.      


\noindent The {\sl Problem of Beables/Observables for RPM's}.  
This paper shows that in the RPM arena, this is a {\sl surmountable} problem.  


\noindent In Isham's words, ``{\it The prime source of the Problem of Time in Quantum Gravity is the invariance of classical GR under the group  
$\mbox{Diff}({\cal S})$ of diffeomorphisms of the spacetime manifold ${\cal S}$}" \cite{I93}.   
This is a central and physically sensible property of GR, and it binds together a lot of the facets of the Problem of Time.     
It would amount to ignoring most of what has been learnt from GR at a fundamental level to simply change to a theory which does not have these complications (such as perturbative string theory on a fixed background).  
Both Loop Quantum Gravity (LQG) \cite{Thiemann} and the more recent M-theory development of String Theory \cite{BerPer} do 
(aim to) take such complications into account.  
The extent to which RPM's are valuable models is to a large extent {\sl bounded} by this difference.
E.g. Isham and Kucha\v{r} \cite{IK85} and much of Isham's review \cite{I84} involve issues not captured by RPM's.  
E.g. 2 + 1 GR \cite{Carlip} and the bosonic string \cite{KK02} embody more of the character of the diffeomorphisms, with midisuperspace bearing an even closer parallel.  
From (\ref{EEE}), it follows that the Frozen Formalism Facet also occurs for RPM's. 
Nevertheless, RPM's manage to be background-independent within their own theoretical setting.  
Additionally, the various RPM counterparts of the Wheeler--DeWitt equation do not exhibit the well-definedness problems of full GR.   

\mbox{ }

\noindent Finally, Halliwell's approach and RPM's form a tight fit as regards Problem of Time facets and modelling a number of aspects of midisuperspace GR (see the Conclusion).

\subsection{Outline of the rest of this paper}

Types of beables and resolution of the Problem of Beables are covered in Sec 2.
\noindent In Sec 3, I summarize and slightly generalize Halliwell's 2003 paper \cite{H03} (also worked upon by him and Thorwart \cite{HT}).
This involves Prob(region in configuration space), entering decoherence functional via a class operator containing 
a window function associated with that region.
Note that this is a timeless quantity and is computed for now within the semiclassical regime.
Moreover, Halliwell's objects O obey Poisson brackets $\{H, \mO\} = 0$, for $H$ the model's analogue of the Hamiltonian constraint.  
This approach has not hitherto been considered for models with nontrivial linear constraints ${\cal L}$in as well. 
This paper does so for some simple RPM's.  
I resolve the Problem of Beables for RPM's in 1- and 2-$d$ in Sec 4.
If $\{{\cal L}\mbox{in}, \mO\} =0$ -- the Kucha\v{r} observables/beables condition -- also holds,  the constructed objects are Dirac 
observables/beables.  
I consider the counterpart of the above work of Halliwell in RPM's, which have linear constraints, in Secs 5 to 7.
As a first route to Halliwell's approach for nontrivial linear constraints/a construction of Dirac beables, I consider the case in which 
\K beables are formally known (Sec 5), exemplifying this with the relational triangle for which this is explicit (Sec 6).
As a second route, in Sec 7 I apply the indirect `$\fG$-act $\fG$-all' method that generalizes both group-averaging and the `best matching' 
method of Barbour \cite{BB82, FileR}.   
I do so firstly for the simple case of sharp-edged window functions of Halliwell 2003 \cite{H03}.  
Paper II will cover these things for the softened version that Halliwell presently advocates \cite{H09}.  
In the Conclusion (Sec 8) I also provide some comments on Halliwell's approach and analyse how well it fares as regards the various facets of the 
Problem of Time and as an implementation of QM propositions; I also consider various possible attitudes to environments in the study of RPM's in Appendix A.

\section{Some notions of observables/beables}

Study of observables in this sense started with Dirac \cite{DiracObs}. 
%
%
\cite{Rovellibook, FileR} include more recent overviews on observables.  
Bell \cite{Speakable, Bell} placed emphasis on conceiving in terms of {\it beables}, which carry no connotations of external observing, but rather 
simply of being, rendering these more suitable for the quantum-cosmological setting.

I also note that it is questionable for the Frozen Formalism Problem to have `primality rights' \cite{APOT2} as regards inducing strategies.  
There are strategies for each and every facet, and then overall strategy is an n-tuple of strategies to face off as many facets at once as possible.  
Thus the `Tempus Ante Quantum, Tempus Post Quantum, Tempus Nihil Est' trichotomy of \cite{Kuchar92, I93} is superseded by a larger set of joint strategies \cite{APOT2}.
In the particular case of the Problem of Observables/Beables, three strategies suggested in the literature are delineated below.

\subsection{Dirac beables}
%

\noindent {\bf Dirac beables} \cite{DiracObs, DeWitt67, +POOref} alias {\bf constants of the motion} alias {\bf perennials} 
\cite{Kuchar93, +Perennials, KucharObs} are any functionals of the canonical variables O = $\mD[\bQ, \bP]$ that, at the classical level, 
have Poisson brackets with all of a theory's constraints $\{{\cal C}_{\sfA}\}$ that vanish (perhaps weakly \cite{I93}), 
\beq
\{ {\cal C}_{\sfA}, \mO\} = 0 \mbox{ } .  
\label{DirObs}
\eeq
Thus, for geometrodynamics  
\beq
\{{\cal H}(x), {\mO}\} =  0 
\mbox{ } ,  \mbox{ } \mbox{ }                  
\{{\cal H}_{\mu}(x), {\mO}\} = 0 \mbox{ } .                     
\label{OHi0}
\eeq
Justification of the name `constants of the motion' conventionally follows from the total Hamiltonian being
%
%
$H\lfloor \Lambda^{\sfA} \rfloor = \int_{\sS}\d \mS \, \Lambda^{\sfA}{\cal C}_{\sfA}$ for multiplier coordinates $\Lambda^{\sfA}$, 
%
%
so that (\ref{DirObs}) implies 
\beq
\d{\mO}[\bQ(x^{\gamma}, t), \bP(x^{\gamma}, t)]/\d t = 0 \mbox{ } . 
\eeq  
The quantum counterpart involves some operator form for the canonical variables and commutators $|[ \mbox{ } , \mbox{ } ]|$ 
in place of $\{ \mbox{ } , \mbox{ } \}$.  
%

\mbox{ } 

\noindent{\bf Alternative Frozen Formalism Facet}: The operator-and-commutator counterparts of the above are then another

\noindent manifestation 
of the Frozen Formalism Problem of classical canonical GR.  
[This can be viewed as some sort of `Heisenberg' counterpart of the `Schr\"{o}dinger' Wheeler--DeWitt equation being frozen.]

\mbox{ }  

\noindent {\bf Kucha\v{r}'s Unicorn} is a sufficient set of Dirac beables to describe one's theory is termed.  
This follows from his quotation ``{\it Perennials in canonical gravity may have the same ontological status as unicorns 
-- a priori, these are possible animals, but a posteriori, they are not roaming on the earth}" \cite{Kuchar93}.  

\mbox{ }  

\noindent Strategy 1) then holds that Dirac beables are conceptually necessary; one needs to know them in order to fully unlock Quantum Gravity.  
In the mythological mnemonic, this makes most sense if one imagines one's Unicorn steed to be winged (like that of the heroine of a well-known 
cartoon from the mid-80's), so as to nullify the Ice Dragon itself having wings.    
  
\mbox{ } 

\noindent The preceding two blocks are based on the following results.  

\mbox{ }  

\noindent {\bf \K 1981 no-go result} \cite{Kuchar81b} is for nonlocal objects of the form  
\beq
\mO = \int_{\sS} \d^3x K_{\mu\nu}(x^{\gamma}; h_{\mu\nu}]\pi^{\mu\nu}(x^{\gamma}) \mbox{ } .
\label{K1981}
\eeq
%

\mbox{ }   

\noindent {\bf Torre 1993 no-go result} \cite{Torre93} (see also \cite{+Torre}) is for local functionals O of $h_{\mu\nu}$ and $\pi^{\mu\nu}$. 

\mbox{ }

\noindent I finally note that for this paper's models, Strategy 1) is implemented via combining the below Strategy 2 
and my generalization of the Halliwell construct.

\subsection{Kucha\v{r} observables/beables} 

Replace (\ref{DirObs}) with split conditions
\beq
\{{\cal Q}\muu\ma\md, \mO\} = 0     \mbox{ } ,
\label{QuadCo}
\eeq
\beq
\{{\cal L}\mi\mn_{\sfZ}, \mO\} = 0  \mbox{ } .   
\label{LinCo}
\eeq

\mbox{ } 

\noindent {\bf Kucha\v{r} beables} \cite{Kuchar93} are as above except that only their brackets with the linear constraints (\ref{LinCo}) need vanish. 
 
\mbox{ } 

\noindent Strategy 2) Kucha\v{r} then argued \cite{Kuchar93} (see also \cite{Kuchar93, KucharObs}) for only the former needing to hold, in which case I denote the objects not by 
O = K$[\bQ, \bP]$ rather than D$[\bQ, \bP]$. 
Thus, here Kucha\v{r} observables are all.  
N.B. it is clear that finding these is a timeless pursuit: it involves configuration space or at most phase space but not the Hamiltonian constraint and thus no dynamics.
The downside now is that there is still a frozen quadratic energy-type quantum constraint on the wavefunctions, 
so that one has to concoct some kind of emergent time or timeless manoeuvre to deal with this.  
The Ice Dragon is here rendered flightless by `disarmament treaty': it `concedes not to have/use its wings' in 
exchange for `a number of one's strategies against it ceasing to be allowed'.  

\mbox{ } 

\noindent Beyond the above-listed literature that this less stringent condition may suffice, 
I also use it below as a technical half-way concept/construct in the formal and actual construction of Dirac beables.  

\mbox{ } 

\noindent As to particular methods for attaining \K beables, 

\mbox{ }

\noindent 1) ${\cal L}$in is associated with some group $\fG$, by which $\mK[\bQ, \bP]$ being $\fG$-invariant suffices for K to be Kucha\v{r}.  
In some cases one can find explicit such constructs (see Sec 4 for some examples); I denote a sufficient set of those by $\mK^{\sfA}$.   
\noindent If one can solve the Best Matching Problem, then this alongside a small amount of canonical workings gives a geometrically-lucid 
solution to the problem of finding the Kucha\v{r} beables.  

\mbox{ }  

\noindent 2) In general, one can however indirectly construct such, at least formally, by taking any object $A$ and applying a $\fG$-act $\fG$-all 
pair to it,\footnote{See \cite{FileR} for a more detailed account of this concept and its scope. 
It is a group sum/group average/group extremization move, which also generalizes Barbour--Bertotti's \cite{BB82} (see also \cite{RWR, FEPI}) 
indirect `best matching' implementation of configurational relationalism.}
e.g. the group action $\stackrel{\rightarrow}{\fG}_g$ followed by integration over the group $\fG$ itself, 
\beq
\mO = \int_{g \in \sFG} \mathbb{D}g \stackrel{\rightarrow}{\fG}_g A \mbox{ } . 
\eeq
This is limited for full quantum gravity by not being more than formally implementable for the case of the 3-diffeomorphisms. 


\noindent A further issue here \cite{v3} is what is the extent of overlap between kinematical quantization's \cite{I84} object selection and 
selection of beables.   
One's classical notion of beable is in each of the above cases to be replaced with the quantum one tied self-adjoint operators obeying a 
suitable commutator algebra in place of the classical Poisson algebra; this correspondence is however nontrivial (e.g. the two algebras 
may not be isomorphic) due to global considerations \cite{I84}.

\subsection{Partial observables/beables}

{\bf True observables} (Rovelli 1991 \cite{Rov91}, see also \cite{Carlip90}) alias {\bf complete observables} (Rovelli 2002 \cite{Rov02}) 
(which at least Thiemann \cite{Thiemann} also calls evolving constant of the motion) classically involve operations on a system each of 
which produces a number that can be predicted if the state of the system is known.    
This conceptualization of observables is closely-related to the above Dirac observables and should then be contrasted with the following much more cleanly distinct conception.

\mbox{ } 
 
\noindent {\bf Partial observables} (Rovelli 1991 \cite{Rov91}) classically involve operation on the system that produces a number that is possibly totally unpredictable even if the state is perfectly known.  
The physics then lies in considering pairs of these objects which between them do encode some extractable purely physical information.  

\mbox{ }  

\noindent Note that while the above definitions were more or less in place by 1991, 
the early 1990's and 2000's forms of the Problem of Time strategies that use these do themselves in part differ.
%
%
\noindent Quantum-mechanically, each of the above two definitions carry over except that the entities whose predictabilities enter the 
definitions are now quantum-mechanical (and the states are now taken to be specifically Heisenberg ones). 

\mbox{ }   

\noindent Strategy 3) In fact it is but partial observables that are necessary.  
\noindent Rovelli and Carlip argued such positions in the early 1990's, with some earlier roots lying in \cite{DeWitt67, PW83}; for subsequent 
such positions, see e.g. \cite{Rov02, Rovellibook, Dittrich}.    
Here the Ice Dragon's ever having possessed wings, and the subsequent need for flighted Unicorns to compensate for this, are held to have always 
been a misunderstanding of the true nature of beables, which are in fact entities that are commonplace but meaningless other than 
{\sl as regards correlations between more than one such considered at once}.  

\mbox{ } 

\noindent 
I include this SSec for completeness and comparison; this paper considers strategies 2 and then 1.

\subsection{Halliwell proto-beables}

Bringing this work of Halliwell's into the discussion at this point is my suggestion.  
This is an alternative to Rovelli's approaches, being rather part of the Dirac--Kucha\v{r} `plot-arc'.  


\noindent {\bf Halliwell proto-beables} are what I term quantities that obey (\ref{LinCo}) and not necessarily (\ref{QuadCo}), 
or possibly the QM counterpart of this statement.  
%
%
\noindent They are therefore the complement of Kucha\v{r} beables, so that 
\beq
\mbox{Kucha\v{r} AND Halliwell } \Rightarrow \mbox{Dirac}.  
\eeq
%
%
Moreover, Halliwell supplies a {\sl specific construct} for a family of these, beginning from\footnote{I use bolds as configuration space vectors, 
and what I present is a slight generalization of Halliwell's presentation, as is suitable for $N$-particle and furtherly relational 
particle models that I move on to cover in subsequent Secs.}
\beq
A(\bq, \bq_0, \bp_0) = \int_{-\infty}^{+\infty} \d t \,  \delta^{(k)}(\bq - \bq^{\sc\sll}(t)) \mbox{ } .  
\label{1stA}
\eeq
It is important to treat the whole path rather than segments of it, for, the endpoints of segments contribute non-negligible right-hand-side 
terms to the attempted commutation with $H$. 
\noindent This and a further implementation at the quantum level in the form of class operators (see below) is conceptually strong: 
its reparametrization-invariance implements temporal relationalism \cite{ARel}, 
the constructs are reasonably universal 
and have a clear meaning in terms of propositions.  
[Though in this last sense, it would appear to require extension at least to its phase space counterpart.]  

\mbox{ } 

\noindent A limitation is that so far Halliwell's approach has only been done  for cases with no linear constraints ${\cal L}\mi\mn_{\sfZ}$.  
Thus the RPM arena is a suitable place in which to further investigate Halliwell's approach, since that has both constraints and a nice amount 
of solvability for the Problem of Beables (Sec 4), though we should first understand the context of Halliwell's work itself, to which we next turn.

\section{Outline of Halliwell 2003 approach}

This approach of Halliwell's \cite{H03} (partly with Thorwart \cite{HT}) involves, in conceptual outline, ``{\it life in an energy eigenstate}".  
%
%
I.e. a timeless approach, albeit framed within Histories Theory, done semiclassically, and, as I further exposit in the present paper, 
with connections to some approaches to the Problem of Beables.

\subsection{Classical preliminary}

Halliwell begins by considering probability distributions, firstly on classical phase space, $\mw(\bq,\bp)$ and then at the semiclassical level.
For the classical analogue of energy eigenstate,
\beq
0 = \frac{\pa \mw}{\pa t} = \{H, \mw\} \mbox{ } , 
\eeq
so $\mw$ is constant along the classical orbits.
Halliwell considers expressions for  
\beq
P_{\sR} := \mbox{Prob}(\mbox{classical solution will pass through a configuration space region R})  \mbox{ } . 
\eeq
This can be taken to be motivated by Mackey's idea \cite{Mackey} that physics should concern propositions about the physical objects and their properties.
This was taken up by Isham and Linden \cite{IL} in the context of Histories Theory and conjectured to be universal to whichever formulation 
of physics/Problem of Time strategy \cite{FileR}.   
See Secs 3.4 and 8.2 in the Conclusion for more about this. 

\mbox{ } 

\noindent Note 1) Such propositions can then be considered {\sl via} probabilities such as the above.

\noindent Note 2) Some propositions require the phase space extension of the above \cite{Omnes2}.

\mbox{ }

\noindent Halliwell evokes $f_{\sR}(\bq)$ as the characteristic function of the region R, and makes use of the phase space function (\ref{1stA}). 
[I use general configuration space and phase space coordinates here, with dim($\fQ$) = $k$ , in contradistinction to his part-implicit 
1-particle in dimension $d$ type notation, due to the $N$-body and relational directions taken in the present article.]
\beq
\mbox{Then }  \mbox{Prob(intersection with R)} = \int_{-\infty}^{+\infty}\d t\,f_{\sR}(\bq^{\sc\sll}(t)) = 
\int \mathbb{D}\bq \, f_{\sR}(\bq) \, \int_{-\infty}^{+\infty}\d t \, \delta^{(k)}\big(\bq - \bq^{\sc\sll}(t)\big) = 
\int \mathbb{D}\bq \, f_{\sR}(\bq) \, A(\bq, \bq_0, \bp_0 )  \mbox{ } ,
\eeq
the `amount of t' the trajectory spends in R.  
Note that I am not just reproducing \cite{H03} here, but also explicitly expressing the totality of the working in terms of a generalized 
coordinate system of a generalized (arbitrary-signature) Riemannian configuration space geometry.   
This is in anticipation of this paper's later workings.  
\beq
\mbox{Then} \hspace{2in}
P_{\sR} = \int \int \mathbb{D}\bp_0 \, \mathbb{D}\bq_0 \, \mw(\bq_0,\bp_0) \,\, \theta
\left(
\int_{-\infty}^{+\infty}\d t\,f_R(\bq^{\sc\sll}(t)) - \epsilon
\right) \mbox{ } .  \hspace{1.8in}
\eeq
Here, the $\btheta${\bf -function} serves to mathematically implement the restriction the entirety of the phase space that is 
being integrated over to that part in which the corresponding classical trajectory spends time $> \epsilon$ in region R. 
$\epsilon$ is some small positive number that tends to 0, included to avoid ambiguities in the $\theta$-function at zero argument.  
Next, as previously stated, the given  $A$ commutes with $H$.

\mbox{ }  

\noindent An alternative expression is for the flux through a piece of a \{$k$ -- 1\}-dimensional hypersurface within the configuration space, 
\beq
P_{\Sigma} = \int\d t\int \mathbb{D}\bp_0 \, \mathbb{D}\bq_0 \, \mw(\bq_0, \bp_0) 
\int_{\Sigma} \mathbb{D}{\Sigma}(\bq) \,\,\, \bn\cdot \bfM \cdot \frac{\d \bq^{\sc\sll}(t)}{\d t} \,\,\, \delta^{(k)}\big(\bq - \bq^{\sc\sll}(t)\big)
= \int \d t \int\mathbb{D}\bp^{\prime} \int_{\Sigma} \mathbb{D}{\Sigma}(\bq^{\prime}) \,\,\, 
  \bn\cdot \bp^{\prime} \,\,\, \mw(\bq^{\prime}, \bp^{\prime}) \mbox{ } , 
\label{33}
\eeq
the latter equality being by passing to $\bq^{\prime} := \bq^{\sc\sll}(t)$ and $\bp^{\prime} := \bp^{\sc\sll}(t)$  coordinates at each $t$.
Here also $\bfM$ is the configuration space metric.  
The first form of (\ref{33}),  
via the cancelling out its of d$t$'s, manifests reparametrization-invariance (and thus temporal relationalism).  

{            \begin{figure}[ht]
\centering
\includegraphics[width=0.3\textwidth]{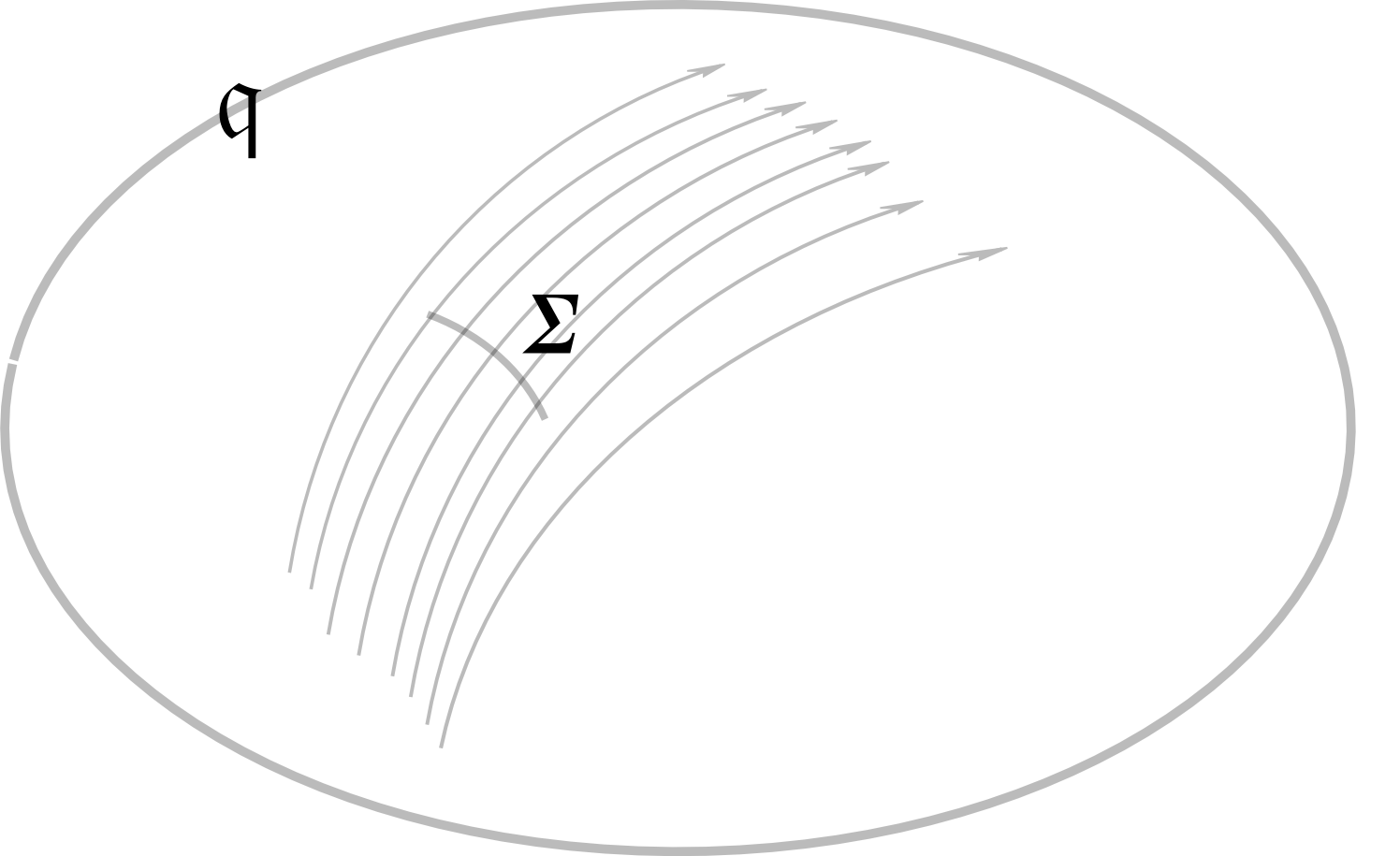}
\caption[Text der im Bilderverzeichnis auftaucht]{        \footnotesize{Halliwell's approach considers fluxes through pieces of hypersurfaces 
$\Sigma$ within configuration space $\fQ$.}    }
\label{Preliminaria} \end{figure}          }

\subsection{Semiclassical working} 

\noindent The last alternative has further parallel \cite{Halliwell87} at the semiclassical level with the {\bf Wigner function} \cite{Wigner}
(see also \cite{HCTBB, B-Bbook}).  
This is a QM phase space distribution function. 
It is but a quasiprobability distribution because it can take negative values; the preceding three references deal in detail with its 
physical interpretation.  
To give the reader a conceptual outline of the Wigner function, a form for it in the flat-configuration-space $K$-dimensional form in the Cartesian case is
\beq
\mbox{Wig}(\bq, \bp) = \frac{1}{\pi^K} \int\int \d^K{\by} \,\langle \psi(\bq + {\by} | \mbox{exp}(2i\,{\by} \cdot \bq ) 
|\psi(\bq - {\by})\rangle  \mbox{ } . 
\eeq
This set-up is such that its integral over $\bp$ gives $|\psi(\bq)|^2$ and its integral over $\bq$ gives $|\psi(\bp)|^2$.  
A distinguishing characteristic of the Wigner function is that its equation of motion is very similar to the classical one; 
in Liouville form, in the Cartesian coordinate case, one has just a correction proportional to (making $\hbar$ explicit)  
$\hbar^2V^{\prime\prime\prime}\pa^3 S/\pa p^3$ for $V$ the potential \cite{B-Bbook}. 


\noindent Semiclassicality helps at this particular point with explicit construction of the class functional in an unambiguous manner 
(this includes \cite{H09, H11}'s modifications).  
Next, starting from the WKB ansatz (\ref{WKB}), Halliwell \cite{Halliwell87} established a further result, 
\beq
\mbox{Wig}(\bq, \bp) \approx |\chi(\bq)|^2\delta^{(K)}(\bp - {\mbox{\boldmath{$\nabla$}}}S) \mbox{ } , 
\label{H87}
\eeq
%
%
with $\bp$ being equal to ${\mbox{\boldmath{$\nabla$}}}S$ at the purely classical level (as per Hamilton--Jacobi theory).   
Then (based on \cite{HP86, Halliwell87, 92})
\beq
P_{\Sigma}^{\sss\se\sm\si\sc\sll} \approx  
\int \d t^{\se\sm}\int_{\Sigma}\mathbb{D}{\Sigma}(\bq) \,\,\, \bn \cdot {\mbox{\boldmath{$\nabla$}}}S \, |\chi(\bq)|^2 \mbox{ } .
\label{36}
\eeq
Halliwell's treatment continues within the standard framework of decoherent histories.
The key step for this continuation is the construction of class operators, which uplifts a number of features of the preceding structures.

\subsection{Class operators}

Class operators, which I denote by $C_{\sR}$, concern Prob(enters a region R of configuration space); $C_{\sR}$ refers to an $\alpha$ comprising of those histories that involves crossing over into region R.\footnote{This is based on the idea of scattering of classical trajectories in the region.  
There are various versions of class operators as per the distinction between Hartle and Halliwell versions, below and \cite{H09, H11}/Paper II's version.
Issues with this are then as follows.  
i)    A slight spreading occurs. 
ii)   The $P_{\tR}$ notion is open to difficulties in general due to chaos.
iii)  Harshness of the $\theta$ function's edges causes a quantum Zeno problem; this is resolved by the construction in \cite{H09} and Paper II.
In fact, via this resolution Prob(does not enter a region R) ends up playing the main role in Halliwell's program.
 
This construct is also related to the Mott \cite{Mott} bubble chamber paradigm.  
This is also alluded to in Barbour's timeless Records Approach \cite{B94IIEOT}, however Mott and most of Barbour are set in space, whilst Halliwell's application resides in configuration space itself.  
Bell \cite{Bell} covers both situations, being the first to combine Mott's idea with Misner's vision of Cosmology \cite{Magic} as scattering in (mini)superspace.}
%
Halliwell's versions \cite{H03, H09, H11} use integrals over time to resolve the issue of compatibility with the Hamiltonian constraint. 

\mbox{ }

\noindent {\bf Mathematical implementation of class operators}.  
One cannot use the most obvious 
\beq
C_{\sR}(\bq_{\sf}, \bq_0) = \int_{-\infty}^{+\infty} \d \lambda\, \mbox{exp}(-i E\lambda) \, 
\int \mathbb{D}\bq(t)\mbox{exp}(iS[\bq(t)]) \, \theta 
\left(   
\int_0^{\lambda} \d t f_{\sR}(\bq(t))  - \epsilon 
\right) 
\eeq
(whose $E$-factor' here comes from \cite{HT} assuming the Rieffel inner product \cite{H03}) because of non-commutation with $H$.  

\mbox{ }

\noindent However, the `sharpened' version,  
\beq
{C}^{{\sharp}}_{\sR} = \theta
\left(   
\int_{-\infty}^{\infty} \d t f_{R}(\bq(t))  - \epsilon 
\right) P(\bq_{\sf}, \bq_0) \,   \mbox{exp}(iA(\bq_{\sf}, \bq_0)) \mbox{ } , \label{gyr}
\eeq
is satisfactory, both conceptually and as regards commutation with $H$.  
The cofactor of $\theta$ above is the standard semiclassical approximation to the unrestricted path integral. 
The nature of the prefactor $P$ is described in \cite{HT} and references therein.  
This is not the end of the story since (\ref{gyr}) is technically unsatisfactory for the reason given in footnote 7, as resolved in \cite{H09} 
(and covered in  Paper II). 
However, the above form does suffice as a conceptual-and-technical start for RPM version of Halliwell-type approaches 
and amounting to an  extension of them to cases including also linear constraints.

\subsection{Decoherence functionals} 

\noindent The decoherence functional is of the form 
\beq
\mbox{Dec}(\alpha, \alpha^{\prime}) = \int_{\alpha}\mathbb{D} \bq \int_{\alpha^{\prime}} \mathbb{D} \bq^{\prime} 
\mbox{exp}(i\{S[\bq(t)] - S[\bq^{\prime}(t)]\}\Rho(\bq_{0}, \bq_{0}^{\prime}) \mbox{ } .
\eeq
\noindent Class operators are then fed into the expression for the decoherence functional:  
\beq
\mbox{Dec}(\alpha, \alpha^{\prime}) = \int\int\int \mathbb{D}\bq_{\sf}\mathbb{D}\bq_{0} \mathbb{D}\bq^{\prime}_{\sf} \, 
{C}^{{\sharp}}_{\alpha}(\bq_{\sf}, \bq_{0}) \, {C}^{{\sharp}}_{\alpha^{\prime}}(\bq^{\prime}_{\sf}, \bq^{\prime}_{0})
\Psi(\bq_{0})  \Psi(\bq^{\prime}_{0}) \mbox{ } .  
\label{dunno}
\eeq
The decoherence functional can be recast in terms of the influence functional \cite{FH} as 
\beq
\mbox{Dec}(\alpha, \alpha^{\prime}) = \int\int\int \mathbb{D} \bq_{\sf} \mathbb{D} \bq_{0} \mathbb{D} \bq_{0}^{\prime}
{C}^{{\sharp}}_{\alpha}(\bq_{\sf}, \bq_0)  {C}^{{\sharp}}_{\alpha}(\bq_{\sf}, \bq_0^{\prime}) {\cal F}(\bq_{\sf}, \bq_0, \bq_0^{\prime})
\Psi(\bq_0)\Psi^*(\bq_0^{\prime}) \mbox{ } .
\eeq 
Then, if \cite{HT}'s conditions hold (involving environment-system interactions), the influence functional 
${\cal F}$ takes the form (in the Cartesian case) 
\beq
{\cal F}(\bq_{\sf}, \bq_0, \bq_0^{\prime}) = \mbox{exp}(i \bq \cdot \bGamma + \bq \cdot \bsigma \cdot \bq)
\eeq
for          ${\bq}^{-} := {\bq} - {\bq}^{\prime}$ and $\Gamma_{\Gamma}$, $\sigma_{\Gamma\Lambda}$ are real coefficients 
depending on ${\bq} + {\bq}^{\prime}$ alone and with $\bsigma$ a non-negative matrix.
Using 

\noindent $\bq^+ := \{\bq_0 + \bq_0^{\prime}\}/2$ as well, the Wigner function is 
\beq
\mbox{Wig}(\bq, \bp) = \frac{1}{\{2\pi\}^K}\int \mathbb{D}\bq \, \mbox{exp}(-i\bp\cdot\bq)\rho(\bq^+ + \bq^-/2, \bq^+ - \bq^-/2)  \mbox{ } . 
\eeq
\beq
\mbox{Then} \hspace{1.7in} P_{\sR} = \int\int \mathbb{D} \bp_0 \mathbb{D}\bq \, \, \theta
\left(
\int_{-\infty}^{+\infty} \d  t^{\se\sm} f_{\sR}\big( \bq^{+\sc\sll}\big) - \epsilon
\right)
\widetilde{\mbox{Wig}}(\bq_0^+, \bp_0)  \hspace{4in}  
\eeq 
for $\bq^{+\sc\sll}(t)$ the classical path with initial data $\bq^+_0, \bp_0$ and Gaussian-smeared Wigner function
%
%
\beq 
\widetilde{\mbox{Wig}}(\bq^+_0, \bp_0) = \int \mathbb{D} \bp \, \mbox{exp}(-\frac{1}{2}
\{\bp_0 - \bp - \Gamma\}\cdot\bsigma\cdot \{\bp_0 - \bp - \bGamma\}) \mbox{Wig} (\bq_0^+, \bp_0)  \mbox{ } .  
\eeq 
 
\noindent Note 1) As regards the various no-go theorems, I note that the above avoids \K 1981 by not being of form (\ref{K1981}) through 
(the field-theoretic generalization of) Halliwell's object being both a $t$-integral and not just linearly-dependent on the momenta 
via e.g. the Wigner function portion of it depending on these in general in more complicated ways.  
It avoids Torre 1993 by not being local in space or time.  

\noindent Note 2) Moreover, by involving a $t$-integral, Halliwell's object is not local in time, which would however be a desirable property in a practically useable observable.  

\noindent Note 3) There may be a region implementation of the propositions problem.  
The ${C}^{\sharp}$ expression's dependence on regions is Boolean: $\mR_1$ OR $\mR_2$ is covered by $f_{R_1 \bigcup R_2}$.  
%
%
However, it is not then clear that this is desirable as regards considering the entirety of the quantum propositions and how these combine, 
as explained in Sec 8.2.

\section{Observables/Beables for RPM's}

\subsection{Characterization of Dirac and \K cases}

For indirectly-formulated RPM, (\ref{DirObs}) takes the form 
\beq
\{{\ttH}, \mO\}       =  0 \mbox{ } ,                     
\label{OE0}  
\eeq
\beq
\{{\ttL}_{\mu}, \mO\} = 0 \mbox{ } ,                     
\label{OLi0}
\eeq
and, also, in the pure-shape case, 
\beq
\{\ttD, \mO\}       =  0 \mbox{ } .                   
\label{OD0}  
\eeq
Then \K observables O = K[$\bQ, \bP$] solve (\ref{OLi0}) for the scaled case, and (\ref{OLi0}, \ref{OD0}) for the pure-shape case.
Dirac observables O = D[$\bQ, \bP$] solve (\ref{OE0}, \ref{OLi0}) for the scaled case and (\ref{OE0}, \ref{OLi0}, \ref{OD0}) 
for the pure-shape case.

This identification of \K observables is a useful application of the results of the program \cite{06I, 06II, TriCl, FORD, AF, 08I, +Tri, ScaleQM, 08III} 
and its antecedents by Kendall \cite{Kendall} and Molecular Physics parallels \cite{LR97}.
This is possible because the Best Matching Problem is solved for 1- and 2-$d$ RPM's (whether pure-shape or scaled) by \cite{FORD, Cones, FileR}.  
%
%
This occurs in pure-shape RPM for precisely the set of all functionals of the shape variables and the shape momenta, $\mK[\bS, \bP_{\sS}]$.
Likewise, the set of  Kucha\v{r} observables for pure-shape RPM is precisely the set of all functionals of the scale and shape variables and the 
scale and shape momenta, $\mK[\sigma, \bS, \mP_{\sigma}, \bP_{\sS}]$.

I can spell out what all of these variables are for pure-shape and scaled RPM's in 1- and 2-$d$. 
\noindent I first need to introduce {\it relative Jacobi coordinates} \cite{Marchal} ${\bR}^i$.  
These are linear combinations of relative position vectors ${\br}^{IJ} = {\bq}^J - {\bq}^I$ between particles into inter-particle cluster vectors such 
that the kinetic term is diagonal.  
Relative Jacobi coordinates have associated particle cluster masses $\mu_i$. 
In fact, it is tidier to use {\it mass-weighted} relative Jacobi coordinates $\brho^i = \sqrt{\mu_i} {\bR}^i$ (Fig \ref{Fig1}). 
The squares of the magnitudes of these are the partial moments of inertia $I^i = \mu_i|{\bR}^i|^2$.  
I also denote $|\brho^i|$ by $\rho^i$ 
%
%
and $\brho^i/\rho$ by $\bn^i$ for $\rho = \sqrt{I}$ the {\it configuration space radius} (alias {\it hyperradius} in the Molecular Physics 
literature \cite{ACG86}).

{            \begin{figure}[ht]
\centering
\includegraphics[width=0.95\textwidth]{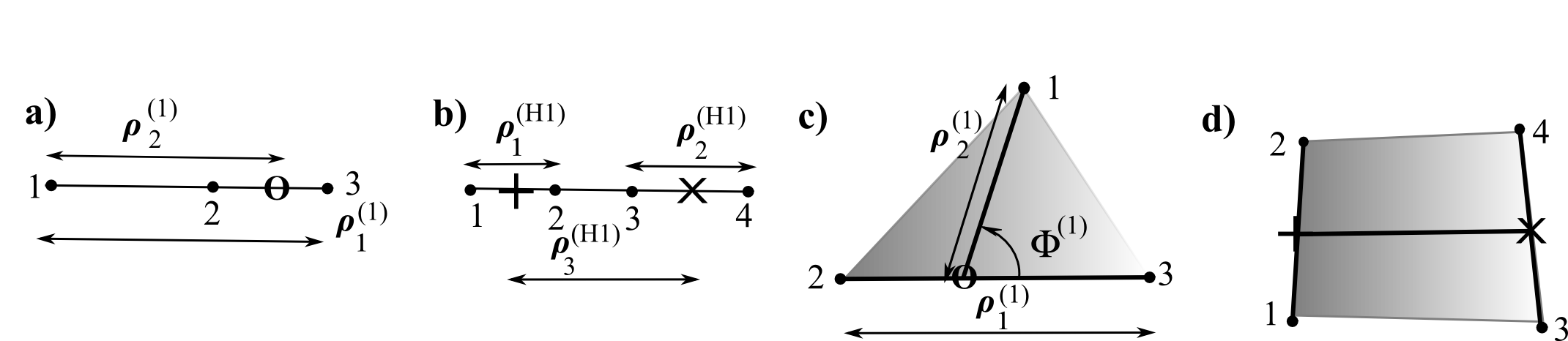}
\caption[Text der im Bilderverzeichnis auftaucht]{        \footnotesize{O, + and $\times$ denote COM(23), 
COM(12) and COM(34) respectively, where COM(ab) is the centre of mass of particles a and b. 
a) is 3-stop metroland and b) is 4-stop metroland in Jacobi H-coordinates.  
\noindent c) is triangleland.
Here, I furthermore define $\Phi^{(\sa)}$ as the `Swiss army knife' angle 
$\mbox{arccos}\big( \brho_1^{(\sa)} \cdot \brho_3^{(\sa)} / \rho_1^{(\sa)} \rho_3^{(\sa)} \big)$.
\noindent d) I provide the 4 particles in the plane layout to make clear what I mean by H-coordinates: these are shaped 
like the strokes in the letter H, which is then `squashed' in the 1-$d$ case.  
The figures' other labels pick out which clustering each coordinate system is with respect to; these are dropped in subsequent computations.}     }
\label{Fig1} \end{figure}          }

The 1-$d$ pure-shape r-configuration spaces are \cite{06II} $\mathbb{S}^{N - 2}$ and suitable shape variables are here the (ultra)spherical angles 
$\Theta^{\barr}$ \cite{AF}, interpreted as functions of ratios of relative separations. 
E.g. for 4-stop metroland (a universe model consisting of 4 particles on a line), these are 
\noindent $\theta = \mbox{arctan}\big(\sqrt{\rho_1\mbox{}^2 + \rho_2\mbox{}^2}/\rho_3\big)$ 
and $\phi = \mbox{arctan}(\rho_2/\rho_1)$.  
The shape momenta are then \cite{AF, FileR, QuadII}
\beq
{\cal D}_{\phi}   := p_{\phi} = \mn_1\mp_2 - \mn_2\mp_1  = {\ttD}_2\mn_1/\mn_2 - {\ttD}_1\mn_2/\mn_1
\label{WillBeA3} \mbox{ and } \mbox{ } 
{\cal D}_{\theta} := p_{\theta} \mbox{ } . 
\eeq
\mbox{ } \mbox{ } The 2 -$d$ pure-shape r-configuration spaces are \cite{Kendall, FORD} $\mathbb{CP}^{N - 2}$ and suitable shape variables 
are here the inhomogenous coordinates ${\mZ}^{\barr}$.  
To interpret these complex coordinates in terms of the $N$-a-gons, it is useful to pass to their polar forms,  
${\mZ}^{\barr} = {\cal R}^{\barr}\mbox{exp}(i\Phi^{\barr})$.  
Then the moduli are, again, ratios of relative separations, and the phases are now relative angles.  
In the specific case of the scalefree triangle, there is one of each, e.g. in coordinates based around the \{1,23\} clustering, 
these are \cite{TriCl} $\Theta = 2\,\mbox{arctan}(\rho_2/\rho_1)$ and $\Phi = \mbox{arccos}\big( \brho_1 \cdot \brho_3 / \rho_1 \rho_3 \big)$ as per Fig 2d).  
The shape momenta for the $N$-a-gon are \cite{FileR, QuadII}
\beq
{\cal P}^{{\cal R}}_{\barp} = 
\left\{
\frac{\delta_{\barp\barq}}{1 + ||{\cal R}||^2}   - 
\frac{{\cal R}_{\barp}{\cal R}_{\barq}}{\{1 + ||{\cal R}||^2\}^2}
\right\}
{\cal R}_{\barq}^{*}
\mbox{ } \mbox{ } , \mbox{ } \mbox{ }  
{\cal P}^{\Theta}_{\widetilde{\sp}} = 
\left\{
\frac{\delta_{\overline{\sp}\overline{\sq}}}{1 + ||{\cal R}||^2} - 
\frac{{\cal R}_{\overline{\sp}}{\cal R}_{\overline{\sq}}}{\{1 + ||{\cal R}||^2\}^2}
\right\}
{\cal R}_{\overline{\sp}}{\cal R}_{\overline{\sq}}\Theta_{\widetilde{\sp}}^{*} \mbox{ } .  
\eeq 
The scalefree triangle subcase \cite{TriCl, FileR, QuadII} can furthermore be expressed in terms of 
${\cal D}_{\triangle} =: p_{\Theta} := \Theta^*$, ${\cal J} =: p_{\Phi} := \mbox{sin}^2\Theta\,\Phi^*$.  
Here, and more generally, I use ${\cal J}$ to denote angular momenta.  
This ${\cal J}$, moreover, clearly cannot be an overall angular momentum since $\ttL = 0$ applies.  
It is indeed a relative angular momentum \cite{08I}: 
\beq
{\cal J} = I_1I_2\Phi^{*}/I = I_1I_2\{\theta_2^* - \theta_1^*\}/I = \{I_1\mL_2 - I_2\mL_1\}/I = \mL_2 = - \mL_1 = \{\mL_2 - \mL_1\}/2 \mbox{ } .
\eeq
Thus it can be interpreted as the angular momentum of one of the two constituent subsystems, minus the 
angular momentum of the other, or half of the difference between the two subsystems' angular momenta.  
That this  is indeed a relative angular momentum is also clear from it being the conjugate of a relative angle.

For scaled RPM's, the shape-scale split \cite{Cones} allows for one just to add a `radial' scale variable to the above sets 
(though there are other presentations too, see below).  
In the polar form that makes the split manifest, they are as for the pure-shape case alongside scale $\rho$ (the hyperradius \cite{LR97} 
which is the square root of the moment of inertia) and the momentum conjugate to this, $p_{\rho} = \rho^*$.  
For the triangle, one needs $I = \rho^2$ to place the angular part into standard spherical form.  
In the Cartesian form for $N$-stop metroland, one has $\rho^i$ coordinates with conjugate momenta $p_i = \rho_j^*$.  
The triangle also admits a Cartesian form: in terms of Dragt-type coordinates \cite{Dragt, +Tri}, 
\beq
\mbox{Dra}_1 = 2\,\brho_1\cdot\brho_2   \mbox{ } , \mbox{ } \mbox{ }  
\mbox{Dra}_2 = 2|\brho_1 \cr \brho_2|_3 \mbox{ } , \mbox{ } \mbox{ } 
\mbox{Dra}_3 =   \brho_1^2 - \brho_2^2    \mbox{ } . 
\eeq
These coordinates can be interpreted \cite{+Tri} as a measure of anisoscelesness aniso, 4 $\times$ the mass-weighted area per unit moment of inertia 
of the triangle, and the ellipticity ellip of the base relative to the median (see \cite{+Tri} for more detail). 
Their conjugates are \cite{FileR, QuadII} 
\beq
\Pi^{\sD\sr\sa}_{\Gamma} = \mbox{Dra}_{\Gamma}^* \mbox{ } 
\eeq
(for $\Gamma$ taking values 1 to 3) which are rates of change of ellipticity (pure-dilational), area and anisoscelesness (these last two are part-dilational and part-rotational).  

{            \begin{figure}[ht]
\centering
\includegraphics[width=0.98\textwidth]{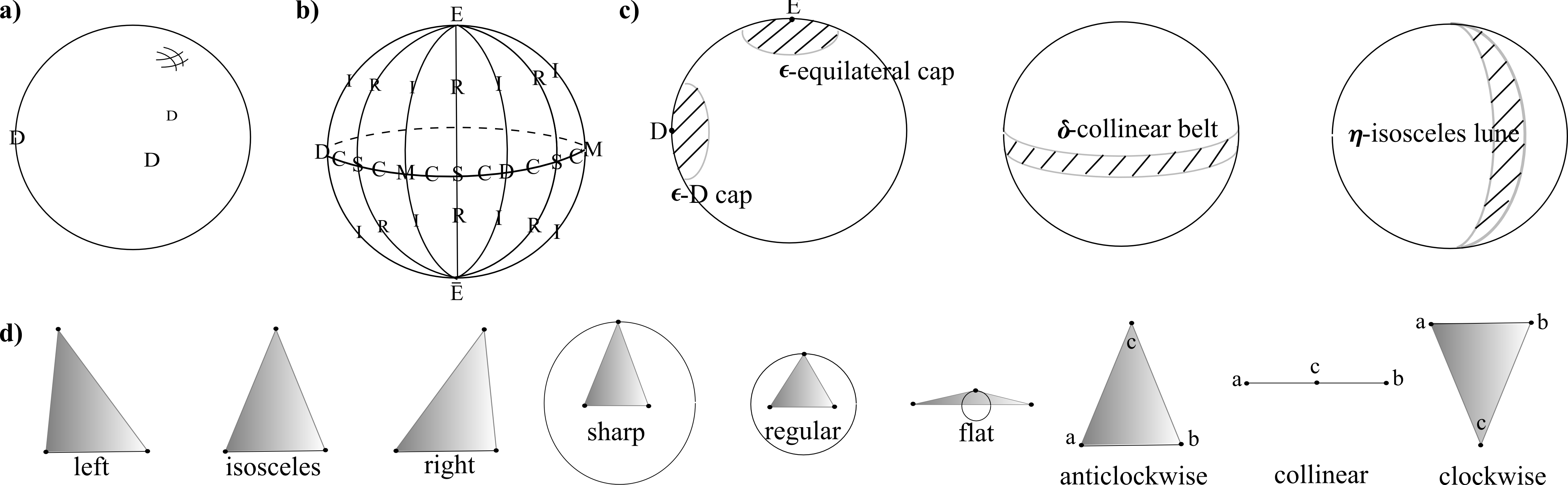}
\caption[Text der im Bilderverzeichnis auftaucht]{        \footnotesize{ a)  Triangleland at the topological level. 
b) Triangleland at the metric level.   
The {\it [ ] basis} has E as principal axis and D as second axis.  
On the other hand, the {\it ( ) basis} has D as principal axis and E as second axis.
c) Some cap, belt and lune regions with particular physical significance \cite{+Tri} are as follows.  
The caps are of radius $\epsilon$, the belts of width $\delta$ and the lunes are of polar angular width $\eta$.  
The above are for pure-shape triangles; the scaled counterparts are in each case just the cones over the pure-shape cases.   
d) The nomenclature I use for dynamically convenient types of triangle. 
The most physically meaningful great circles on the triangleland shape correspond to the isosceles, regular and collinear triangles.
These respectively divide the shape sphere into hemispheres of right and left triangles, sharp and flat 
triangles, and anticlockwise and clockwise triangles.     }   }
\label{Triada} \end{figure}          }

I give more detail of triangleland here, since scaled triangleland is the specific example that this paper makes use of.
The configuration space for this in the case of distinguishable particles and in the plain shape case is the sphere, decorated 
as in Figs \ref{Triada}b) and \ref{Triada}f).
The labelled points and edges have the following geometrical/mechanical interpretations.  
E and $\bar{\mE}$ are the two mirror images of labelled equilateral triangles.  
C are arcs of the equator that is made up of collinear configurations. 
This splits the triangleland shape sphere into two hemispheres of opposite orientation (clockwise and anticlockwise labelled 
triangles, as in Fig \ref{Triada}c). 
The I are bimeridians of isoscelesness with respect to the 3 possible clusterings (i.e. choices of base pair and apex particle).   
Each of these separates the triangleland shape sphere into hemispheres of right and left slanting triangles with respect 
to that choice of clustering [Fig \ref{Triada}c)].
The R are bimeridians of regularness (equality of the 2 partial moments of inertia of the each of the possible 
2 constituent subsystems: base pair and apex particle.)
Each of these separated the triangleland shape sphere into hemispheres of sharp and flat triangles with respect to that 
choice of clustering [Fig \ref{Triada}d)].  
The M are {\it merger points}: where one particle lies at the centre of mass of the other two. 
S denotes spurious points, which lie at the intersection of R and C but have no further notable properties (unlike the D, M or E points 
that lie on the other intersections).

It is sometimes also convenient to swap the Dra$_{2}$ for the scale variable $I$ in the non-normalized version 
of the coordinates to obtain the \{$I$, Dra$_1$, Dra$_3$\} system.  
Then a simple linear recombination of this is \{$I_1$, $I_2$, Dra$_1$\}, i.e. the 
two partial moments of inertia and the anisoscelesness.  
This is in turn closely related \cite{08I} to the parabolic coordinates on the flat $\mathbb{R}^3$ conformal to the 
triangleland relational space, which are $\{I_1, I_2, \Phi\}$.

A physical interpretation for these is that $I_1$, $I_2$ are the partial moments of inertia of the base and the median, with $\Phi$ the `Swiss 
army knife' angle between these (c.f. Fig \ref{Fig1}) . 
They are clearly a sort of subsystem-split coordinates and thus useful in applications concerning subsystems \cite{FileR, ARel}.

In each case, the quantum counterpart involves some operator form for the canonical variables and commutators in place of Poisson brackets.  
E.g. in the configuration representation the constituent operator variables are the shapes again, alongside differential operators for the 
shape momenta.  
In the case of the scaled triangle, these have the mathematics of the standard angular momentum operators albeit now with 
a more general interpretation as relative shape momenta (mixed relative-dilational and relative-angular momenta),
\beq
\mbox{sin}\,\Phi\, {\cal D}_{\triangle} + \mbox{cos}\,\Phi\,\mbox{cot}\,\Theta\,{\cal J}   \mbox{ } \mbox{ and } 
\mbox{ } \mbox{ }
-\mbox{cos}\,\Phi\, {\cal D}_{\triangle} + \mbox{sin}\,\Phi\,\mbox{cot}\,\Theta\,{\cal J}   \mbox{ } .
\eeq
\noindent There is the further issue use of conserved quantities in preference to/alongside the momenta.  

\noindent 1) These, or functions thereof, commute also with the Hamiltonian constraint and are thus Dirac beables.  
They manage this via not encountering an obstruction from the potential in the bracket of O and $H$.  

\noindent 2) They feature in the kinematical quantization procedure, making them even more natural at the quantum level.  
For the sphere, these are the $SU(2)$ quantities ${\cal S}_i$.  
Here also e.g. for the sphere, $\Phi$ and $\Theta$ are not good operators, it is, rather, the unit Cartesian vectors whose 
squares sum up to 1 that are.  
In total, one has ${\cal S}_i$, Dra$^{\Gamma}$ and $\Pi^{\sD\sr\sa}_{\Gamma}$, forming the algebra 

\noindent Eucl(3) $\mbox{\textcircled{S}} \mathbb{R}^3$, for Eucl(3) the `Euclidean group' of translations and rotations (of 3-$d$ reduced 
configuration space) and where \textcircled{S} denotes semi-direct product.  



The next simplest example concerns shape quantities and theory conjugate momenta for the quadrilateral and is presented in 
\cite{QuadI, QuadII}, which are built on the kinematical considerations of \cite{QShapeQSub} 



As regards higher-$d$ RPM's \K observables, firstly I emphasize that these models are not needed to toy-model geometrodynamics \cite{FileR}.  
Secondly, here the Best Matching Problem has a global obstruction to solvability -- one can only invert the higher-$d$ inertia tensor present 
in the ${\cal L}$-constraint if one excludes the collinear configurations.  
And yet, these configurations are entirely physical so this localness of procedure is phys unsatisfactory.  
See Appendix 3.E of \cite{FileR} for further detail of this second point.

\subsection{Dirac observables solved for: Strategy 2 resolved}

\noindent If one instead adheres to needing the more restrictive subset of complete observables, then one is to ask 
which functions of shape and shape momentum commute with the pure-shape RPM quadratic energy constraint, 
and which functions of  scale, shape, scale momentum and shape momentum commute with the scaled RPM  quadratic energy constraint.  



A simple partial answer is that in a few cases these include (subsets) of the isometries, i.e. relative angular momenta, relative 
dilational momenta and linear combinations of these with certain shape-valued coefficients.  
E.g. this is a direct analogue of the angular momenta forming a complete set of commuting operators with the Hamiltonian operator, provided that the potential is central i.e. itself respecting the isometries of the sphere by being purely radial. 
Thus for the tower of $SO(p)$-symmetric problems (2 $\leq p \leq n$) for $N$-stop metrolands and triangleland, and its more elaborate counterpart for 
quadrilateralland in \cite{QuadII}, we have found some complete observables.  
These are not, however, generically present i.e. for arbitrary-potential models. 
Another answer, at least in some simple models is that Halliwell's class operators comply with ${\cal Q}\mbox{uad}$.   
For at least some simple $\fG$-trivial theories, Halliwell's class operators are Dirac observables, and they or their 
phase space generalization may provide the complete set of such.  
See the next 3 Secs for more about this.

\section{One way to extend Halliwell's work to constrained theories}

\subsection{Nature of the extensions}

\noindent Extension 1) I consider the case of linearly-constrained theories that are at least formally reducible. 
My treatment here 

\noindent is purely quadratic/bosonic for simplicity.
$\bK$ is a coordinate vector for configurational \K beables; these are independent and there are the right number 
$r$ = dim($\fQ^{\sr}$) of them to span.  
The complete set of \K beables are more general: functionals $F[\bK,\bPi^{\tK}]$.
Having \K beables explicitly available has close ties with one's model  being reducible; both are in practise exceptional circumstances. 
They do however apply to 1- and 2-$d$ RPM's, which are indeed both reducible and have full sets of \K beables known (c.f. Sec 6's example).  
We present an alternative strategy in Sec 7 which, whilst indirect, is more widely applicable in cases without reducibility or knowledge 
of an explicit directly-expressed set  of \K beables.  
Full use of \K beables would involve the extension of Halliwell's construct to such as phase space regions since these are general functionals of  
$\bQ$, and $\bP$ rather than just of $\bQ$. 

\mbox{ } 

\noindent Additionally, I consider a relational whole-universe context for which the following are held to apply.

\mbox{ }

\noindent Extension 2) Emergent time $t^{\se\sm}$ emerges to fill in the role of $t$; this emergent time is the coincidence of \cite{SemiclI}
Jacobi--Barbour--Bertotti time \cite{BB82, B94I} at the classical level and WKB time at the semiclassical level \cite{HallHaw}.  

\noindent Extension 3) Parageodesic principle conformal transformation (PPCT) invariance is held to apply by  Misner's argument \cite{Magic} 
for conformal invariance as the second selector within DeWitt's family of configuration space 
recoordinatization-invariant operator orderings \cite{DeWitt57} being grounded in the classical relational whole-universe models' action 
\cite{Banal, FileR} being held to continue to apply at the quantum level.  
This involves the relational action (\ref{Actio}) being manifestly invariant under the {\sl internal} conformal invariance
\beq
\d s_{\sr}^2 \rightarrow \d\overline{s}^2 = \Omega^2\d s_{\sr}^2 \mbox{ } , \mbox{ } \mbox{ } 
 E - V \rightarrow \{\overline{E} - \overline{V}\} = \{E - V\}/\Omega^2  \mbox{ }   
\label{PPCT}
\eeq
(my notation here restricting to the reduced subcase, but this also applying to the minisuperspace GR setting of Misner).   
I next note that this implies the inverse of the configuration space metric to scale as 

\noindent
\beq
N_{\sfA\sfB} \rightarrow \overline{N}_{\sfA\sfB} = \Omega^{-2}N_{\sfA\sfB}
\label{PPCTinv}
\eeq 
and the square-root of the determinant of the configuration space metric to scale as 
\beq
\sqrt{|M|} \rightarrow \sqrt{|\overline{M}|} = \Omega^r \sqrt{|M|} \mbox{ } .  
\label{PPCTdet}
\eeq  
One next recovers Misner's conformal covariance of the Hamiltonian constraint (or its generalization, ${\cal Q}$uad). 
Then taking this to carry over to the quantum level alongside DeWitt's configuration space recoordinatization invariance implies that 


\noindent i) the conformal operator ordering (which the preceding identifies as originating from specifically PPCT-invariance), 
\beq
N^{\sfA\sfB}_{\sr}P^{\sK}_{\sfA}P^{\sK}_{\sfB} 
\mbox{ } \mbox{ } 
\stackrel{\mbox{\scriptsize promote to a quantum operator}}{\longrightarrow} 
\mbox{ } \mbox{ } 
\triangle^{\sc}_{\sr} := \triangle_{\sr} - \frac{r - 2}{4\{r - 1\}}\mbox{Ric}(M^{\sr}) \mbox{ } , 
\eeq 
\beq
\mbox{for} \hspace{2.4in}
\triangle := \frac{1}{\sqrt{M^{\sr}}}\frac{\pa}{\pa \mK^{\sfA}}
\left\{ 
\sqrt{M^{\sr}}N_{\sr}^{\sfA\sfB}\frac{\pa}{\pa \mK^{\sfB}}
\right\}    \hspace{3.3in}
\eeq
the usual Laplacian corresponding to $M^{\sr}_{\sfA\sfB}$.  


\noindent ii) The wavefunctions are also then to scale as (see e.g. \cite{Wald})
\beq
\Psi \longrightarrow \overline{\Psi} = \Omega^{\{2 - r\}/2}\Psi \mbox{ } .  
\label{PPCTwave}
\eeq  
\noindent iii) For the physical quantities to be invariant, the corresponding inner product needs a weight function $\omega$ 
PPCT-scaling as 
\beq
\omega \longrightarrow \overline{\omega} = \Omega^{-2}\omega 
\label{PPCTweight}
\eeq 
(see e.g. \cite{08II}), so that, by use of (\ref{PPCTdet}, \ref{PPCTwave}, \ref{PPCTweight})  
\beq
\int \overline{\mathbb{D}\bK \, \Psi_1^*\Psi_2 \omega} = 
\int \mathbb{D}\bK \Omega^r  \, \Psi_1^*\Psi_2 \Omega^{2\{2 - r\}/2} \omega \Omega^{-2} = 
\int \mathbb{D}\bK \, \Psi_1^*\Psi_2 \omega  \mbox{ } .  
\eeq


\noindent Note that having explicit \K observables implies reducibility, so that there is then no formal 
barrier to performing the above PPCT-invariant interpretation of conformally-invariant ordering.  
This matters insofar as conformally-invariant ordering does not in general commute with applying linear constraints \cite{FileR}, 
 thus jeopardizing the argument for this ordering in all those case in which one cannot explicitly reduce first. 


\noindent iv) Another ready corollary of (\ref{PPCT})  \cite{Banal} is that the emergent time element scales as 
\beq
\d t^{\se\sm} \longrightarrow \d \overline{t}^{\se\sm} = \Omega^2\d t^{\se\sm} \mbox{ } .  
\eeq
\noindent It then follows as a new result of this paper that $f_{\sR}$ 
needs to scale as $\Omega^{-2}$ so that the overall combination $\d t f_{\sR}$ is PPCT-invariant.  
\noindent The below are also all new to this paper.  


\noindent v) If one applies a PPCT to an $r$-manifold with metric $M_{\sfa\sfb}$ containing an \{$r$ -- 1\}-dimensional hypersurface with metric $m_{\sfa\sfb}$ and normal $n^{\sfa}_{\sK}$, then the formula for the induced metric implies that 
\beq
m_{\sfa\sfb} \longrightarrow \overline{m}_{\sfa\sfb} = \Omega^2m_{\sfa\sfb} 
\mbox{ }  \mbox{ and } \mbox{ } 
n_{\sK \sfa} \longrightarrow \overline{n}_{\sK \sfa} = \Omega n_{\sK \sfa}
\eeq
from which it immediately follows that 

\noindent
\beq
\sqrt{|m|} \longrightarrow \overline{\sqrt{|m|}} = \Omega^{r - 1}\sqrt{m} 
\mbox{ } \mbox{ and } \mbox{ } 
n_{\sK}^{\sfa} \longrightarrow \overline{n}_{\sK}^{\sfa} = \Omega^{-1}n_{\sK}^{\sfa} \mbox{ } .  
\eeq
\noindent vi) If (\ref{PPCTwave}) applies to a wavefunction obeying the BO-WKB ans\"{a}tze form (\ref{BO}, \ref{WKB}), 
then preservation of the physically-significant h--l split under PPCT transformations requires it to be entirely the $\chi($\mh$, $\ml$)$ factor that PPCT-scales, 
\beq
\chi \longrightarrow \overline{\chi} = \Omega^{\{2 - r\}/2}\chi
\eeq
since $\Omega$ itself is in general a function of h and l, and so would map $S$(h) out of the functions of h alone.
Likewise it is the inner product integrating over the l-coordinates that carries a $\Omega^{-2}$ factor.


\noindent vii) The outer rather than inner product of two wavefunctions necessitates the same weight function; 
this will of course be used to build density matrices.  


\noindent viii) The phase space measure does not PPCT-scale, as a result of the momentum space measure scaling oppositely
to the configuration space one.  


\noindent ix) Finally, I posit that the classical probability density $\mw$ is PPCT-invariant, so that 
$\int \mathbb{D}\bK\,\mathbb{D}\bPi^{\sK} \mw(\bK, \bPi^{\sK})$ is also.

\subsection{Classical preliminary} 

Parallelling Halliwell, I begin by considering probability distributions, firstly on classical phase space, $\mw(\bK,\bPi^{\sK})$ and then at the semiclassical level.
For the classical analogue of energy eigenstate,
\beq
0 = \frac{\pa \mw}{\pa t^{\se\sm}} = \{H, \mw\} \mbox{ } ,
\eeq
so $\mw$ is constant along the classical orbits.
I evoke $f_{\sR}(\bq)$ as the characteristic function of the region $\mR$, and makes use of a phase space function $A$ that is now not just 
any $A$ but an $A$ based on Kucha\v{r} observables $\bK$: 
\beq
A(\bK, \bK_0, \bPi^{\sK}_0) = \int_{-\infty}^{+\infty} \d t^{\se\sm} \delta^{(r)}(\bK - \bK^{\sc\sll}(t^{\se\sm}) ) \mbox{ } . 
\label{candi}
\eeq
\beq
\mbox{Then} \hspace{1.8in} 
\{{\cal L}\mi\mn, A(\bK, \bK_0, \bPi^{\sK}_0)\} = 
\int_{-\infty}^{+\infty} \d t^{\se\sm} \{{\cal L} \mi\mn, \delta^{(r)}(\bK - \bK^{\sc\sll}(t^{\se\sm}) )\} \mbox{ } ,
\label{qirk} \hspace{2.2in} 
\eeq
\beq
\mbox{and} \hspace{3.15in} 
\{ {\cal L}\mi\mn, \bK\} = 0 \hspace{3.4in} 
\eeq
because the $\bK$ are Kucha\v{r}, so $\{ {\cal L}\mi\mn, f(\bK)\} = 0$ by the chain-rule, and so 
\beq
\{{\cal L}\mi\mn, A(\bK, \bK_0, \bPi^{\sK}_0)\} = 0  \mbox{ } .
\label{uni2}
\eeq
Moreover, being in terms of a vector of Kucha\v{r} observables does not change the argument by which 
\beq
\{ {\cal Q}\muu\ma\md, A(\bK, \bK_0, \bPi^{\sK}_0)\} = 0 
\label{corn2}
\eeq
(which Halliwell has already demonstrated to be robust to curved configuration space use). 
Thus, by (\ref{uni2},\ref{corn2})  combined, 


\noindent 1) (\ref{candi}) are Dirac.


\noindent 2) As substantial a set of Dirac observables can be built thus for a theory whose full set of Kucha\v{r} observables are known 
as could be built for Halliwell's simpler non linearly-constrained theories.  
To that extent, one has a formal construction of the Unicorn. 
[Though, at the QM level, this role of A is played out again by the class operators instead.]  

\mbox{ } 

\noindent Moreover, in the case of 1- and 2-$d$ RPM's, scaled or unscaled, Sec 4 ensures that this is an {\sl actual} construction for these 
toy models' toy Unicorn.

\noindent 
$$
\mbox{Next,} \hspace{0.5in} 
\mbox{Prob(intersection with R)} = \int_{-\infty}^{+\infty}\d t^{\se\sm}\,f_{\sR}(\mbox{\bf K}^{\sc\sll}(t^{\se\sm})) = 
\int \mathbb{D}\mbox{\bf K} \, f_{\sR}(\bq) \, \int_{-\infty}^{+\infty}\d t^{\se\sm} \delta^{(r)}
\big(\mbox{\bf K} - \mbox{\bf K}^{\sc\sll}(t^{\se\sm})\big) \hspace{4in} 
$$
\beq
= \int \mathbb{D}\mbox{\bf K} \, f_{\sR}(\bK) \, A(\mbox{\bf K}, \mbox{\bf K}_0, \bPi^{\sK}_0 ) :
\eeq
the `amount of $t^{\se\sm}$' the trajectory spends in R; moreover this physical quantity is constructed to be PPCT-invariant by iv).

\noindent 
\beq
\mbox{Then,} \hspace{1.2in} 
P_{\sR} = \int \mathbb{D}\bPi^{\sK}_0 \, \mathbb{D}\mbox{\bf K}_0 \, \mw\big(\mbox{\bf K}_0,\bPi^{\sK}_0\big) \,\, \theta
\left(
\int_{-\infty}^{+\infty}\d t^{\se\sm} f_{\sR}\big(\mbox{\bf K}^{\sc\sll}(t^{\se\sm})\big) - \epsilon
\right) \mbox{ } , \hspace{4in} 
\eeq
which is PPCT-invariant through coming in three factors each of which is PPCT-invariant [by viii) and ix)].  

\mbox{ } 

\noindent Note that the window function corresponding to the region R is assumed to fit on a single coordinate system, limiting it to being somewhat local.  
This is entirely fine if one is considering small regions (see Sec 6 for more).  
This also continues to work approximately for compact RCS's like pure-shape RPM examples have (c.f. Fig \ref{Triada}).  


An alternative expression is for the flux through a piece of an \{$r$ -- 1\}-$d$ hypersurface within the configuration space, 
$$
P_{\Sigma} = \int\d t^{\se\sm}\int\int \mathbb{D}\bPi^{\sK}_0 \, \mathbb{D}\mbox{\bf K}_0 \, \mw(\mbox{\bf K}_0,\bPi^{\sK}_0) 
\int_{\Sigma} \d^2{\Sigma}(\mbox{\bf K}^{\prime}) \, \bn^{\sK\prime} \cdot \frac{\d \mbox{\bf K}^{\sc\sll}(t^{\se\sm})}{\d t^{\se\sm}} \, \delta^{(r)}\big(\mbox{\bf K} - \mbox{\bf K}^{\sc\sll}(t^{\se\sm})\big)
$$
\beq
= \int \d t^{\se\sm} \int\mathbb{D}\bPi^{\sK \prime} \int_{\Sigma} \mathbb{D}{\Sigma}(\mbox{\bf K}^{\prime})\, \bn^{\sK\prime}\cdot\bPi^{\sK\prime} \, \mw(\mbox{\bf K}^{\prime}, \bPi^{\sK}) \mbox{ } , 
\eeq
the latter equality being by passing to $\mbox{\bf K}^{\prime} := \mbox{\bf K}^{\sc\sll}(t^{\se\sm})$ and $\bPi^{\sK} := \bPi^{\sK}_{\sc\sll}(t^{\se\sm})$ 
coordinates at each $t^{\se\sm}$.
This is again PPCT-invariant, by v), viii) and ix).

\subsection{Semiclassical quantum working}

\noindent Extension 4): Wigner function in curved space. 
As well as previous considerations of volume elements, this has the further subtlety that the sums inside the bra and ket are no longer trivially 
defined.  
This was resolved by Winter, Calzetta, Habib, Hu and Kandrup \cite{WKCHH}, by Fonarev \cite{Fon} and by Liu and Qian \cite{LQ} in the case of 
Riemannian configuration space geometry via local geodesic constructions.
%
%

\noindent Underhill's earlier study \cite{Underhill} works with just affine structure assumed.
(In searching for this topic in the literature, it is useful to note that the Wigner function is closely related to the Weyl transformation; 
see also Sec 2.3 of the review \cite{Landsman}.)     
Liu and Qian also extended their work \cite{LQ} to principal bundles over Riemannian manifolds, thus covering what is required to extend Sec 7 in 
terms of Wigner functions.
Because of this, I specifically take `Wigner functions in curved space' in the sense of Liu and Qian when in need of sufficiently detailed 
considerations.
Finally, I emphasize again that Wigner functions are only temporary passengers in the present program due to their being used in Halliwell 2003 
types implementations of class operators but no longer in Halliwell 2009 ones, by which I keep the account of this SSec's subtleties brief.   

\mbox{ }  

\noindent The preceding alternative expression has further parallel with the Wigner function at the semiclassical level.  
Next, \cite{Halliwell87}'s straightforward approximations in deriving (\ref{H87}) locally carry over, so 
\beq
\mbox{Wig}(\mbox{\bf K}, \bPi^{\sK}) \approx |\chi(\mbox{\bf K})|^2\delta^{(r)}(\bPi^{\sK} - {\mbox{\boldmath{$\nabla$}}}^{\sK}S)\label{Wig}
\eeq
($\bPi^{\sK}$ being ${\mbox{\boldmath{$\nabla$}}}^{\sK} S$ for classical trajectories). 
Then Halliwell'-type heuristic move is then to replace $w$ by Wig in (\ref{clvers3}), giving
\beq
P_{\Sigma}^{\sss\se\sm\si\sc\sll} \approx \int\d t^{\se\sm}\int_{\Sigma}\d\Sigma(\bK) \bn^{\sK} \frac{\nabla S}{\nabla\mbox{\bf K}} 
|\chi(\mbox{\bf K})|^2 \mbox{ } .  
\eeq
This remains PPCT-invariant as the quantum inner product and the classical $\int \d\bPi^{\sK} \mw(\bK, \bPi^{\sK})$ both scale  
equally as $\Omega^{-r}$.

\subsection{Class operators}

The Halliwell-type treatment continues within the framework of decoherent histories, which I take as formally standard for this setting too.  
The key step for this continuation is the construction of class operators, which uplifts a number of features of the preceding structures.
One now uses  
\beq
{C}^{\sharp}_{\sR} = \theta
\left(   
\int_{-\infty}^{+\infty} \d t f_{\sR}(\bK(t))  - \epsilon 
\right) P(\bK_{\sf}, \bK_0) \,   \mbox{exp}(iA(\bK_{\sf}, \bK_0)) \mbox{ } , \label{spoo}
\eeq
which, by construction, obeys
\beq
|[ {\cal Q}\mbox{uad}, C^{\sharp}_{\sR} ]| = 0  \mbox{ } ;
\eeq
\beq
|[ {\cal L}\mbox{in}, C^{\sharp}_{\sR} ]| = 0 
\eeq
also holds as  $C^{\sharp}_{\sR}$ is a functional of \K beables.
As the cofactor of $\theta$ is some approximand to the quantum wavefunction, it PPCT-scales as $\Omega^{\{2 - r\}/2}$.

Again, this class operator is not the end of the story since it is technically unsatisfactory, as resolved in \cite{H09, H11} (and to be covered in Paper II), but the above form serves as a conceptual-and-technical start for RPM version of the work and extension to cases with linear constraints, and for the present conceptual, whole-universe and linear-constraint extending paper, this is as far as we shall go.

\subsection{Decoherence functional}

\noindent The decoherence functional is of the form 

\noindent
\beq
\mbox{Dec}(\alpha, \alpha^{\prime}) = \int_{\alpha}\mathbb{D} \bK \int_{\alpha^{\prime}} \mathbb{D} \bK^{\prime} 
\mbox{exp}(i\{S[\bK(t^{\se\sm})] - S[\bK^{\prime}(t^{\se\sm})]\}\Rho(\bK_{0}, \bK_{0}^{\prime})\omega_{\sd} \mbox{ } .
\eeq
\noindent For this to be PPCT-invariant as befits a physical quantity, it needs to have its own weight $\omega_{\sd}$, PPCT-scaling as 
$\omega_{\sd} \longrightarrow \overline{\omega}_{\sd} = \Omega^{-r}\omega_{\sd}$.
Class operators are then fed into the expression for the decoherence functional, giving 

\noindent
\beq
\mbox{Dec}(\alpha, \alpha^{\prime}) = \int\int\int \d^3\bK_{\sf}\d^3\bK_{0} \d^3\bK^{\prime}_{\sf}C^{{\sharp}}_{\alpha}(\bK_{\sf}, \bK_{0}) C^{{\sharp}}_{\alpha^{\prime}}(\bK^{\prime}_{\sf}, \bK^{\prime}_{0})\Psi(\bK_{0})  \Psi(\bK^{\prime}_{0})\omega^2\omega_{\sd} \mbox{ } .  
\label{dunno2}
\eeq 
\noindent The $\omega^2$ factor has one $\omega$ arise from the density matrix and the other from the 2-wavefunction approximand expressions from the two $C^{\prime}$'s.
If the universe contains a classically-negligible but QM-non-negligible environment as per Appendix A, the influence functional 
${\cal F}$ makes conceptual sense and one can rearrange (\ref{dunno2}) in terms of this into the form  
\beq
\mbox{Dec}(\alpha, \alpha^{\prime}) = \int\int\int \mathbb{D}\bK_{\sf} \mathbb{D}\bK_{0} \mathbb{D}\bK_{0}^{'}
C_{\alpha}^{\sharp}(\bK_{\sf}, \bK_0)  C_{\alpha}^{\sharp}(\bK_{\sf}, \bK_0)^{'} {\cal F}(\bK_{\sf}, \bK_0, \bK_0^{'})\Psi(\bK_0)\Psi^*(\bK_0^{'})\omega^2\omega_{\sd} \mbox{ } .  
\label{83}
\eeq

\section{Example: r-presentation of triangleland}


\subsection{Classical counterpart}

Regions of configuration space for RPM's, includes cases of particularly lucid physical significance as per Sec 4's tessellation interpretation.  
Now,
\beq
P_{\sR} := \mbox{Prob}(\mbox{classical solution will pass through a triangleland configuration space region R}) \mbox{ } . 
\eeq 
Next, I evoke $f_{\sR}(\mbox{\bf Dra})$ as the characteristic function of the triangleland configuration space region R, and makes use of phase space functions
\beq
A(\mbox{\bf Dra}, \mbox{\bf Dra}_0, \bPi^{\sD\sr\sa}_0) = \int_{-\infty}^{+\infty}\d t^{\se\sm} \delta^{(3)}
\big(
\mbox{\bf Dra} - \mbox{\bf Dra}^{\sc\sll}(t^{\se\sm})
\big) \mbox{ } .  
\eeq
These commute with ${\cal L}$ and ${\cal E}$ by the argument around equations (\ref{qirk}-\ref{corn2}).  
Then  
$$
\mbox{Prob(intersection with R)} = \int_{-\infty}^{+\infty}\d t^{\se\sm}\,f_{\sR}(\mbox{\bf Dra}^{\sc\sll}(t^{\se\sm})) = 
\int \mathbb{D} \mbox{\bf Dra} \, 
f_{\sR}(\bq) \, \int_{-\infty}^{+\infty}\d t^{\se\sm} 
\delta^{(k)}\big(\mbox{\bf Dra} - \mbox{\bf Dra}^{\sc\sll}(t^{\se\sm})\big) 
$$
\beq
= \int \mathbb{D} \mbox{\bf Dra} \, f_{\sR}(\mbox{\bf Dra}) \, A(\mbox{\bf Dra}, \mbox{\bf Dra}_0, \bPi^{\sD\sr\sa}_0 ) \mbox{ } :
\eeq
the `amount of $t^{\se\sm}$' the trajectory spends in region R.  
Then 
\beq
P_{\sR} = \int \d^3\bPi^{\sD\sr\sa}_0 \, \d^3\mbox{\bf Dra}_0 \, \mw\big(\mbox{\bf Dra}_0,\bPi^{\sD\sr\sa}_0\big) \,\, \theta
\left(
\int_{-\infty}^{+\infty}\d t^{\se\sm} f_{\sR}\big(\mbox{\bf Dra}^{\sc\sll}(t^{\se\sm})\big) - \epsilon
\right)   \mbox{ } .  
\eeq
Halliwell illustrated this with a free particle model; this has a counterpart for the r-presentation of the scaled triangle free classical solution via the {\it Dragt correspondence} 
of \cite{Cones, FileR} which amounts to transcribing Halliwell's mathematics to an arena in which it has whole-universe significance.

\mbox{ }

\noindent An alternative expression is for the flux through a piece of a 2-$d$ hypersurface within the configuration space, 
$$
P_{\Sigma} = \int\d t^{\se\sm}\int \d^3\bPi^{\sD\sr\sa}_0 \, \d^3\mbox{\bf Dra}_0 \, \mw(\mbox{\bf Dra}_0, \bPi^{\sD\sr\sa}_0) 
\int_{\Sigma} \d^2{\Sigma}(\mbox{\bf Dra}) \, \bn^{\sD\sr\sa} \cdot \bfM \cdot \frac{\d \mbox{\bf Dra}^{\sc\sll}(t^{\se\sm})}{\d t^{\se\sm}} \, \delta^{(k)}\big(\mbox{\bf Dra} - \mbox{\bf Dra}^{\sc\sll}(t^{\se\sm})\big)
$$
\beq
= \int \d t^{\se\sm} \int\mathbb{D}\bPi^{\sD\sr\sa \prime} \int_{\Sigma} \mathbb{D}{\Sigma}(\mbox{\bf Dra}^{\prime})\, \bn^{\sD\sr\sa\prime} 
\cdot \bPi^{\sD\sr\sa\prime} \, \mw(\mbox{\bf Dra}^{\prime}, \bPi^{\sD\sr\sa\prime}) \mbox{ } , 
\eeq
the latter equality being by passing to $\mbox{\bf Dra}^{\prime} := \mbox{\bf Dra}^{\sc\sll}(t^{\se\sm})$ and $\bPi^{\sD\sr\sa\prime} := \bPi^{\sD\sr\sa}_{\sc\sll}(t^{\se\sm})$ coordinates at each $t^{\se\sm}$.

\subsection{Semiclassical quantum working}\label{Wigi}

The last alternative above further parallel at the semiclassical level with the Wigner function. 
Now, including a power of the PPCT conformal factor,
\beq
\mbox{Wig}(\bPi^{\sD\sr\sa}, \mbox{\bf Dra}) \approx |\chi(\mbox{\bf Dra})|^2\delta^{(3)}(\bPi^{\sD\sr\sa} - {\mbox{\boldmath{$\nabla$}}}^{\sD\sr\sa}S)/\{4\mI\}^{3/2} \mbox{ } \label{Wig2}
\eeq
($\bPi^{\sD\sr\sa} = {\mbox{\boldmath{$\nabla$}}}^{\sD\sr\sa}S$ for classical trajectories).
I note that in my setting of interest, $|\chi\rangle = |\chi(\Theta, \Phi, \mI)\rangle$ and $S = S(\mI)$. 
%
%
Halliwell's heuristic move is then to replace $\mw$ by Wig in (\ref{clvers3})
\beq
P_{\Sigma}^{\sss\se\sm\si\sc\sll} \approx \int \d t^{\se\sm}\int_{\Sigma}\mathbb{D}\Sigma(\mbox{Dra}) \bn^{\sD\sr\sa} \frac{\nabla S}{\nabla\mbox{\bf Dra}} 
|\chi(\mbox{\bf Dra})|^2 \mbox{ } . \label{lallo}  
\eeq
\noindent The RPM case of most interest is that with the radial `scale of the universe' direction having particular h-significance, by which the configurational 2-surface element is a piece of sphere with a number of these carrying lucid significance by Sec 4 and the 3-momentum 3-surface element is the spherical polars one (modulo conformal factors).
Moreover, the $S$ makes the evaluation of this in spherical polars natural, even if $\Sigma$ itself is unaligned with those 
(though it simplifies the calculation if 

\noindent there is such an alignment).

So rewriting (\ref{lallo}) in conformal--spherical polar coordinates, e.g. for 
\beq
\mbox{\scriptsize Prob(universe attains size I$_0$ $\pm \delta$I whilst being $\epsilon$-equilateral)} \approx  
\int_{t^{\se\sm} = t_0 - \delta t}^{t_0 + \delta t} \d t^{\se\sm} 
\int_{\Phi = 0}^{2\pi}\int_{\Theta = 0}^{\epsilon} 
\mI^2 \mbox{sin}\,\Theta\,\d\,\Theta\d\,\Phi 
\frac{\d S(\mI)}{\d \mI}
|\chi(\mI, \Theta, \Phi)|^2 \mbox{ } .  
\eeq
This made use of this question by addressed by the $\epsilon-cap$ about the E-pole, which is very simply parametrized by the coordinates in use 
[see Fig \ref{Triada}c) for this cap and the below belt].  
Also, $S = S(\mI)$ alone, so $\bn^{\sD\sr\sa\prime} {\nabla S}/{\nabla\mbox{\bf Dra}^{\prime}}$ becomes a radial $\bn^{\sI}\,\pa S/\pa\mI$ factor and two zero components. 
%
%
To proceed, $\d t^{\se\sm} \frac{\d S}{\d \sI}$ = $\d t^{\se\sm} \frac{\d \sI}{\d t^{\se\sm}} = \d \mI$ by the Hamilton--Jacobi expression for momentum, 
the momentum-velocity relation and the chain-rule, so we do not need to explicitly evaluate $t_0 \pm \delta t$ in terms of $\mI \pm \delta I$.
Then e.g. for the approximate semiclassical wavefunction from the explicit triangleland example in \cite{SemiclIII} 
(the upside-down harmonic oscillator for the universe at zero energy), 
\beq
\mbox{\scriptsize{Prob(universe attains size I $\pm \delta$I whilst being $\epsilon$-equilateral)}} \approx  
2 \mI_0^2 \delta\mI   
\int_{\Theta = 0}^{\epsilon}\int_{\Phi = 0}^{2\pi}   |Y_{\sS\,\ms}(\Theta, \Phi)|^2\mbox{sin}\,\Theta\,\d\Theta\, \d\Phi + O(\delta \mI^2)
\label{clvers3} \mbox{ } .
\eeq
Here, the $Y_{\sS\sss}$ are spherical harmonics indexed by triangleland's total shape momentum in the [\mbox{ }] basis defined in Fig \ref{Triada}'s caption.  
The answer then comes out with leading term proportional to e.g. $\mI_0^2\delta\mI \, \epsilon^2$ for all the axially-symmetric wavefuntions
$\chi_{\sS\,0} \propto Y_{\sS\,0}$ and to $\mI_0^2\delta\mI\,\epsilon^4$ for the first non-axial wavefunctions (the sine and cosine 
combinations corresponding to the quantum numbers S and $|\ms|$ = 1).  
These answers make good sense as regards the axisymmetric wavefuntions being peaked around the equilateral triangle whilst the equilateral 
triangle is nodal for the first non-axisymmetric wavefunctions.  

\mbox{ }  

\noindent Note 1) This is an \NSII-type construct, though it is for the semiclassical l-part, so there is some kind 
of semiclassical imprint left on it.

\noindent Note 2) {Prob}(universe attains size I $\pm \delta$I whilst being $\epsilon$-D) is given likewise but for a particular D being given 
by the same in the corresponding ( ) basis, 
and the words ``one orientation" or ``a particular D" being suppressible by summing over various such integrals.  
 
\noindent Note 3) As a final example,  
\beq
\mbox{\scriptsize Prob(universe attains size I $\pm \delta$I whilst being $\epsilon$-collinear)}  \mbox{ } 
\mbox{ has $\epsilon$-cap replaced by $\epsilon$-belt about the equator in the [ ] basis} \mbox{ } ,
\eeq
and the answer goes as as $\mI_0^2\delta\mI\,\epsilon^3$ for the odd-$\mS$ axisymmetric wavefunctions and as $\mI_0^2\delta\mI\,\epsilon$ for the even-$\mS$ axisymmetric 
wavefunctions and the first non-axisymmetric ones.   
This has one power of $\epsilon$ less than for the above example since cap area $\propto \epsilon^2$ but belt area $\propto \epsilon$ only.
The extra $\epsilon^2$ factor can again be explained in terms of peaks and nodes: a nodal plane of collinearity as compared to peaks on 
all or part of it.

\subsection{Class operators}

Again, one uses the modified version, which here takes the form  
\beq
{C}_{\sR}^{\sharp} = \theta
\left(   
\int_{-\infty}^{+\infty} \d t f_{\sR}(\mbox{\bf Dra}(t^{\se\sm}))  - \epsilon 
\right) P(\mbox{\bf Dra}_{\sf}, \mbox{\bf Dra}_0) \,   \mbox{exp}(iA(\mbox{\bf Dra}_{\sf}, \mbox{\bf Dra}_0)) \mbox{ }  .  
\eeq
\beq
\mbox{This obeys} \hspace{2.1in} |[{\cal E}, C^{\sharp}]| = 0 \mbox{ } , \mbox{ } \mbox{ } |[{\cal L}, C^{\sharp}]| = 0 \mbox{ } . \hspace{2.5in}
\eeq

\subsection{Decoherence functionals}

Class operators are then fed into the expression for the decoherence functional,   
\beq
\mbox{Dec}(\alpha, \alpha^{\prime}) = \int_{\alpha}\d^3\mbox{\bf Dra} \int_{\alpha^{\prime}} \d^3\mbox{\bf Dra}^{\prime} 
\mbox{exp}(i\{S[\mbox{\bf Dra}(t^{\se\sm})] - S[\mbox{\bf Dra}(t^{\se\sm})])\}\Rho(\mbox{\bf Dra}_{\si\sn}, \mbox{\bf Dra}_{\si\sn}^{\prime}) 
\mbox{ } ,
\eeq
\beq
\mbox{giving} \hspace{0.6in}
\mbox{Dec}(\alpha, \alpha^{\prime}) = \int\int\int \d^3\mbox{\bf Dra}_{\sf}\d^3\mbox{\bf Dra}_{0} \d^3\mbox{\bf Dra}^{\prime}_{\sf} \, 
{C}^{\sharp}_{\alpha}(\mbox{\bf Dra}_{\sf}, \mbox{\bf Dra}_{0}) \, {C}^{\sharp}_{\alpha^{\prime}}(\mbox{\bf Dra}^{\prime}_{\sf}, 
\mbox{\bf Dra}^{\prime}_{0})\Psi(\mbox{\bf Dra}_{\si\sn})  \Psi(\mbox{\bf Dra}^{\prime}_{\si\sn}) \mbox{ } . 
\label{dunno3} \hspace{0.6in}
\eeq
Under classically insignificant, QM significant environment assumption under which the influence functional is justified,\footnote{Note 
however that the eventual target of paralleling \cite{H09} differs in not requiring environments, at least for `larger regions', 
so not having an alternative at this stage is not a long-term hindrance to the present program.  
It is more a case of \cite{H03} coming with environment-based reservations (not optimal for a fully closed system study) as well as a quantum Zeno problem, 
\cite{H09} but both of these issues go away upon passing to the more advanced \cite{H09} construction in Paper II.
If there is no environment, we lose (\ref{pipupi}), and if \cite{HT}'s justification fails we lose eqs (\ref{snarl}--\ref{whiffle}).}  
\beq
\mbox{Dec}(\alpha, \alpha^{\prime}) = \int\int\int \d^3 \mbox{\bf Dra}_{\sf} \d^3 \mbox{\bf Dra}_{0} \d^3 \mbox{\bf Dra}_{0}^{\prime} \, 
{C}^{\sharp}_{\alpha}(\mbox{\bf Dra}_{\sf}, \mbox{\bf Dra}_0) \, {C}^{\sharp}_{\alpha}(\mbox{\bf Dra}_{\sf}, \mbox{\bf Dra}_0^{\prime}) 
{\cal F}(\mbox{\bf Dra}_{\sf}, \mbox{\bf Dra}_0, \mbox{\bf Dra}_0^{\prime})\Psi(\mbox{\bf Dra}_0)\Psi^*(\mbox{\bf Dra}_0^{\prime}) \mbox{ } .  
\label{pipupi}
\eeq 
\noindent Then if \cite{HT}'s conditions apply [which they do according to Attitude 3) of Appendix A], 

\noindent   
\beq
{\cal F}(\mbox{\bf Dra}_{\sf}, \mbox{\bf Dra}_0, \mbox{\bf Dra}_0^{\prime}) = \mbox{exp}(i\,\mbox{\bf {Dra}} \cdot \bGamma + 
\mbox{\bf Dra} \cdot \bsigma \cdot \mbox{\bf Dra}/2  )  \mbox{ } .  
\label{snarl}
\eeq
If the above step holds, then the below makes sense too. 
Here,          $\mbox{\bf Dra}^{-} := \mbox{\bf Dra} - \mbox{\bf Dra}^{\prime}$ and 
$\Gamma_{\Lambda}$, $\sigma_{\Gamma\Lambda}$ real coefficients depending on $\mbox{\bf Dra} + {\bf Dra}^{\prime}$ alone and with $\bsigma$ 
a non-negative matrix.
Using $\mbox{\bf Dra}^+ := \{\mbox{\bf Dra}_0 + \mbox{\bf Dra}_0^{\prime}\}/2$ as well, the Wigner function is 
\beq
\mbox{Wig}(\mbox{\bf Dra}, \bPi^{\sD\sr\sa}) = 
\frac{1}{\{2\pi\}^3}\int \d^3\mbox{\bf Dra} \, \mbox{exp}(-i\bPi^{\sD\sr\sa}\cdot\mbox{\bf Dra})\rho(\mbox{\bf Dra}^+ + \mbox{\bf Dra}^-/2, 
\mbox{\bf Dra}^+ - \mbox{\bf Dra}^-/2) \mbox{ } . 
\eeq
\beq
\mbox{Then} \hspace{0.7in}
P_{\sR} = \int\int \d^3 \bPi^{\sD\sr\sa}_0\d^3\mbox{\bf Dra} \, \, \theta
\left(
\int_{-\infty}^{+\infty} \d  t^{\se\sm} f_{\sR}\big( \mbox{\bf Dra}^{+\sc\sll}\big) - \epsilon
\right)
\widetilde{\mbox{Wig}}(\mbox{\bf Dra}_0^+, \mbox{\bf Dra}_0)  \hspace{3in}
\eeq 
for $\mbox{\bf Dra}^{+\sc\sll}(t^{\se\sm})$ the classical path with initial data $\mbox{\bf Dra}^+_0, \bPi^{\sD\sr\sa}_0$ and Gaussian-smeared 
Wigner function
\beq 
\widetilde{\mbox{Wig}}(\mbox{\bf Dra}^+_0, \bPi^{\sD\sr\sa}_0) = \int \d^3\bPi^{\sD\sr\sa} \, \mbox{exp}(-\frac{1}{2}
\{\bPi^{\sD\sr\sa}_0 - \bPi^{\sD\sr\sa} - \bGamma\} 
\cdot \bsigma \cdot   
\{\bPi^{\sD\sr\sa}_0 - \bPi^{\sD\sr\sa} - \bGamma\})  
\mbox{Wig}(\mbox{\bf Dra}_0^+, \bPi^{\sD\sr\sa}_0) \mbox{ } .  
\label{whiffle}
\eeq 
This final step is the one in which Halliwell's setting gives a good classical recovery with a smeared Wigner function 
in place of a classical probability distribution.  
%

\section{Alternative indirect $\fG$-act, $\fG$-all extension}

\beq
\mbox{Here,}  \hspace{1.5in}
{A}^{{\sharp}\sfg-\sf\sr\se\se}(\brho, \bp_{0}, \brho_{0}) = \int_{g \in \sfG}\mathbb{D} g \, \stackrel{\rightarrow}{\fG_g}
\int_{-\infty}^{+\infty} \d t^{\se\sm} \, \delta^{(k)}\big( \brho -\brho^{\sc\sll}(t^{\se\sm})\big) \hspace{3in}  
\eeq
\beq
\mbox{and} \hspace{.4in} 
{C}^{{\sharp}\sfg-\sf\sr\se\se}(\brho_{\sf}, \brho_{0}) = \int_{g \in \sFG}\mathbb{D} g \, \stackrel{\rightarrow}{\fG_g}
\left\{
\theta
\left(
\int_{-\infty}^{+\infty} \d t^{\se\sm} \, f_{\sR}\big(\brho_0^{\sf}(t^{\se\sm})) - \epsilon
\right)
P(\brho_{\sf}, \brho_0) \mbox{exp}(i A(\brho_f, \brho_0)
\right\} \mbox{ } . \hspace{3.3in}
\eeq
It is indeed physically desirable for these to already be individually $\fG$-invariant.
Then making the decoherence functional out of ${C}^{\sharp\sfg-\sf\sr\se\se}_{\sR}$ (and noting there is an issue of then needing to average multiple times, though at least $\fG$-averaging a $\fG$-average has no further effect, making this procedure somewhat less ambiguous than it would have been otherwise),
$$
D(\alpha, \alpha^{\prime}) = \int_{g \in \sFG}\mathbb{D} g \, \stackrel{\rightarrow}{\fG_g} \left\{
\int\int\int \mathbb{D}\bq_{\sf}\,\mathbb{D}\bq_0\,\mathbb{D}\bq_0^{\prime}\, 
{{C}}^{{\sharp}\sfg-\sf\sr\se\se}_{\alpha}(\bq_{\sf}, \bq_0) {C}^{{\sharp}\sfg-\sf\sr\se\se}_{\alpha^{\prime}}(\bq_{\sf}, \bq_0^{\prime})\Psi(\bq_0)\Psi(\bq_0^{\prime})\right\} 
$$
\beq
= \int_{g \in \sFG}\mathbb{D} g \, \stackrel{\rightarrow}{\fG_g} \left\{
\int\int\int \mathbb{D}\bq_{\sf}\,\mathbb{D}\bq_0\,\mathbb{D}\bq_0^{\prime}\, 
{{C}}^{{\sharp}\sfg-\sf\sr\se\se}_{\alpha}(\bq_{\sf}, \bq_0) {{C}}^{{\sharp}\sfg-\sf\sr\se\se}_{\alpha^{\prime}}(\bq_{\sf}, \bq_0^{\prime})
{\cal F}(\bq_{\sf}, \bq_0, \bq_0^{\prime})\Psi(\bq_0)\Psi(\bq_0^{\prime})\right\}  \mbox{ } .  
\eeq

\mbox{ }  

\noindent 
For the triangleland example, $\fG = SO(2) = U(1)$, so 
\beq
\int_{g \in \sFG}\mathbb{D} g = \int_{\zeta \in \mathbb{S}^1} \mathbb{D} \mathbb{\zeta} = \int_{\zeta = 0}^{2\pi}\d\zeta
\eeq
for $\stackrel{\rightarrow}{\fG_g}$ the action of the infinitesimal 2-$d$ rotation matrix $\underline{\underline{R}}\mbox{}_{\zeta}$ 
on the vectors of the model, and $\zeta$ the absolute rotation.  

\mbox{ } 

\noindent 
A problem with this alternative approach is that it becomes blocked early on as regards more-than-formality for the case of the 3-diffeomorphisms.

\section{Conclusion}

\subsection{Summary of results so far} 

In this paper, the Problem of \K Observables/Beables is solved for RPM's.
These are functions of the shapes \cite{FORD, FileR, QuadI} (and scale in the scaled RPM) alongside their conjugates the shape momenta \cite{QuadII, FileR}.
Secondly, I extend this to a resolution of the Problem of Dirac Observables for RPM's by use of the class functionals of Halliwell 2003 \cite{H03}, 
which commute with the quadratic constraint as well.  
This also amounts to extending Halliwell's 2003 approach (combined Histories, Records, Semiclassical approach) \cite{H03} for Quantum Cosmology 
to models exhibiting all of nontrivial linear constraints, nontrivial structure formation/inhomogeneity along the lines of the Halliwell--Hawking 
midisuperspace approach \cite{HallHaw} and whole-universe effects.  
Whole-universe effects exhibited in this paper include the universe possessing an emergent time and a conformal invariance (which is the 
same as Misner's \cite{Magic} but now anchored to the relational form of the action \cite{Banal, FileR}). 
See \cite{DeWitt67, FileR} for more closed-universe effects exhibited by RPM's. 
I exemplify the above extension with the concrete example of the relational triangle, using the nice control permitted by the explicitly available 
and simple \K beables available in this case.  
Other theories for which \K beables are known include theories for which the notion is trivial (e.g. minisuperspace), 
a few midisuperspaces such as the cylindrical wave \cite{K72}, spherically symmetry \cite{K94} and some Gowdy models \cite{TorreGowdy}.  
%
%
\noindent I also consider the case in which \K beables are held not to be available, by the indirect $\fG$-act, $\fG$-all method.  
This has wider scope albeit it allows for less formal progress in the general case.
\noindent Passing from classical \K beables to QM ones requires choice of a subalgebra of them that are to be promoted to QM operators.

\subsection{Problem of Time position sum-up for Halliwell-type approaches}

\noindent In addition to the case for this given in the Introduction, I note that 1) each of the histories, records and semiclassical approaches to be combined
can be individually studied in the RPM arena (which qualitatively models two midisuperspace features). 
Each of these strategies has some shortcomings, but the three of them together remove a number of each others' shortcomings.

\noindent 2) Using the RPM arena allows one to operate free of the Foliation Dependence Problem, Functional Evolution Problem, Spacetime Reconstruction Problem 
and Inner Product Problem.

\noindent 3) Using the 1- or 2-$d$ RPM arena has the Best Matching Problem generalization of the Sandwich Problem is resolved 
for them \cite{TriCl, FORD, Cones}.
This allows for the emergent JBB time to be known explicitly, and this is a classical resolution of the FFP and also allows for \K observables to be known.  
This can be uplifted to a construct for Dirac observables at the classical level following Halliwell.  


\noindent 4) The classical emergent JBB time resolution breaks down at the QM level, but can be replaced by the emergent WKB time resolution, 
itself needing justification.
Combined schemes such as Halliwell's go toward that.  

\noindent 5) Thus the combination of Halliwell's approach and RPM's gets round six of the eight facets of the POT.\footnote{It is for this purpose that 
I developed this area at the level of understanding concrete models' configuration spaces, classical dynamics and QM  
\cite{TriCl, FORD, 08I, 08II, AF, +Tri, Cones, ScaleQM, 08III, FileR, QuadI, QuadII}.}

\mbox{ }  

\noindent Caveat 1) Halliwell's approach does not deal with the Global POT or the Multiple Choice Problem, both of which still occur 
for RPM's; the Global POT is harder to deal with at QM level.   
Some further idea would be needed here, perhaps along the lines of Bojowald et al's {\bf fashionables} approach \cite{Bojo}, or my {\bf  degradeables} parallel \cite{FileR}. 
[Fashionables are observables local in time and space, whereas degradeables are beables that are local in time and space.  
These are good words for local concepts, viz `fashionable in Italy', `fashionable in the 1960's', `degradeable outside of the freezer' 
and `degradeable within a year' all making good sense. 
Also, fashion is in the eye of the beholder -- observer-tied, whereas degradeability is a mere matter of being, rather than of any observing.] 
%
%
These are {\bf patching approaches}: observables/beables, and timefunctions, are held to only be valid on certain local patches.  
In the classical case, one can consider this in terms of coordinate charts with limited ranges of validity; however at the quantum 
level one is faced with the open question of how to formulate and interpret an analogue of patching for multiple unitary evolutions.  
Moreover, I point out that the conceptual content of patching/fashionables/degradeables apply not only within Rovelli-type partial observables 
strategies, but also in programs with other attitudes to observables/beables, in particular to \K and Dirac beables based approaches.    
At the level of \K observables for RPM's, using fashionables/degradeables means that the functionals of the shapes (and scales) and their 
conjugates do not need to be valid over the whole of configuration space.
Thus a classical resolution of Problem of \K beables, that complies with the Global Problem of Time and complies at least somewhat 
more with the Multiple Choice Problem, is in terms of local functions of the shapes that obey suitable differential geometry meshing conditions.  
The QM part of this (and patching approaches more generally) remains to be resolved; I have however found an arena -- the RPM's -- in which to investigate the spatial 
as well as hitherto considered temporal aspects of the problem.
I note furthermore that the class functional (\ref{spoo}) is indeed a locally interpretable concept: select a region, so using such a construct to solve the Problem of (Dirac) 
Observables/Beables at the quantum level is indeed compatible with the basic ethos of fashionables/degradeables.  


\noindent Caveat 2) Moreover, by involving a $t$-integral, Halliwell's object is not local in time, which would however be a desirable property in 
a practically useable observable.

 
\noindent Caveat 3) Whilst Halliwell's scheme does not a priori use a na\"{\i}ve Schr\"{o}dinger inner product, it still does use a na\"{\i}ve
Schr\"{o}dinger-type implementation of propositions by configuration space regions (or maybe a generalization to phase space regions).  
This suffers from probabilities corresponding to regions composing too simply (Booleanly) for one to be able to represent all QM propositions in such a form.  
(E.g. quantum propositions may be taken to obey quantum logic \cite{QLog, IL2}, which is nondistributive, unlike the composition of 
configuration space regions, which is distributive [and elsewise Boolean].)  
This is immediately obvious for the \NSI \cite{FileR}, but requires further considerations to establish in the case of a Halliwell-type program, 
due to how having probability 1 for a slice of a flux tube implies probability 1 in other slices of the tube.  
I procede here by taking sections of configuration space; on these, the Booleanness of region composition is as clear as for 
the \NSII.  
As Fig \ref{FigLast} explains, such sections may not always be available, even for composing regions with flow-evolutions of other regions along tubes, 
but they are available enough to be able to have counter-examples (cases in which QM propositions do not obey a Boolean algebra but a section {\sl is} 
available so as to carry out composition of probabilities within Halliwell's scheme).  


{            \begin{figure}[ht]
\centering
\includegraphics[width=1.0\textwidth]{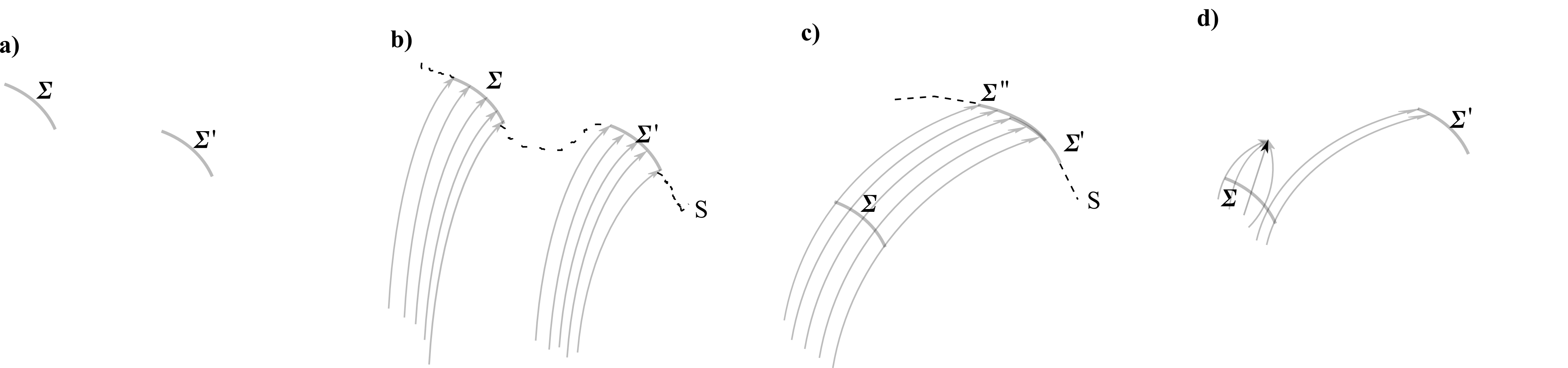}
\caption[Text der im Bilderverzeichnis auftaucht]{        \footnotesize{a) Consider now two hypersurface elements $\Sigma$ and $\Sigma^{\prime}$.
b) Sometimes there will exist a section S orthogonal to the classical flow that includes both $\Sigma$ and $\Sigma^{\prime}$.  
Then via (\ref{33}) or (\ref{36}), P$_{\Sigma\,\,\,\sO\sR\,\,\, \Sigma^{\prime}}$ = P$_{\Sigma\,\bigcup\, \Sigma^{\prime}}$ just as for the \NSII.   
c) However, if $\Sigma$ and $\Sigma^{\prime}$ do not lie on the same section, it is not directly like in the \NSII. 
Now one can consider the flow-evolution of region $\Sigma$, say, so that it lies in a section that extends $\Sigma^{\prime}$; one can now compose 
$\Sigma^{\prime}$ and $\Sigma^{\prime\prime}$ just like one composed in case a). 
d) Moreover there need not always exist a section that extends $\Sigma^{\prime}$ whilst also containing a flow evolution image 
$\Sigma^{\prime\prime}$ of $\Sigma$. 
The construction of counterexamples, however, can avoid this case. 
What this case represents is that the composition of Halliwell's implementation is not always reducible to a parallel of the \NSII's composition.
Sometimes local sections and meshing conditions between them suffice for composition, but this does not always hold either. 
These non-existences reflect that some flows can be pathological/exhibit breakdowns in well-definedness or smoothness.}    }
\label{FigLast} \end{figure}          }


\noindent One way out of Caveat 3) is that QM may require modifying such as to obey an intuitionistic logic \cite{BI, ID}, which {\sl is}
distributive. 
I also note that regions are conceptually different in this kind of approach (they are now primary themselves rather than `made up of' 
some notion of points that are held to be primary), so that this possibility lies outside the scope of the present paper. 
It is also worth mentioning that [in e.g. \cite{PW83, IL, FileR}] proposition--projector association makes better sense than evoking regions, at the quantum level.
This covers e.g. the case of the Isham--Linden approach to Histories Theory (\cite{IL2, IL, ILConcat} see also \cite{Flori}), the \CPI [c.f. (\ref{CPI})], 
or Records within Histories theory \cite{GMH, H99}. 
The main atemporal part of Isham and Doering \cite{ID} could also be viewed an underlying structure for Records Theory.

\subsection{Halliwell 2009 type extension}  

\noindent Extending Halliwell's 2009 work to the RPM arena mostly concerns defining class operators somewhat differently so as to get these to 
be better-behaved as regards the quantum Zeno effect.  
In detail, \cite{H03} amonnts to using an impenetrable barrier potential whilst \cite{H09, H11} corresponds to a slightly penetrable barrier potential. 
This amounts to the region in question being taken to contain a potential, with the class operator being the corresponding S-matrix 
and the slightly penetrable case representing a softening in the usual sense of scattering theory (albeit in configuration space rather than in space).  
[The smoothed-out case manages to avoid the quantum Zeno effect in addition to managing to still be compatible with the (quadratic, not for now linear) constraint.]
Similar mathematics arises in the arrival time problem of QM \cite{ATOverlap, AT}.

\subsection{Quadrilateralland extension}

Scaled $N$-stop metrolands RPM's straightforwardly lie already within the Halliwell 2003 scheme.  
The quadrilateralland counterpart of the present paper is rather more complicated and may be worked out once quantization, the Semiclassical Approach 
and Histories Theory have been considered for that model.

\subsection{More minimalist alternatives}

\noindent The Gambini--Porto--Pullin approach \cite{GPP} is a competitor insofar as it is a timeless/semiclassical combination 
(their work with Torterolo \cite{GPPT} furthermore combines this with ideas about observables).   
Thus it is similarly motivated to Halliwell's approach and similarly applicable to Problem of Time and other issues in the foundations of Quantum Cosmology.  
%
%
It can also be investigated in the RPM arena.
 Arce \cite{Arce} has provided a distinct combined Conditional Probabilities Interpretation--semiclassical scheme.

\subsection{Conceptual analogy between RPM shapes and LQG knots} 

There is a loose conceptual analogy between pure-shape RPM's shapes or scaled RPM's scale-and-shapes (classical resolutions of the linear constraints) 
and LQG's knot states \cite{PullinGambini} (quantum resolutions of the linear constraints). 
Loops themselves are analogous to preshapes [i.e. the configurations prior to the main part of the reduction -- Dil($d$) but not Rot($d$) 
taken out for RPM's or $SU(2)$ but not Diff($\Sigma$) taken out for LQG].  
Passing to knot equivalence classes of loops is then analogous to the transformations used in \cite{FORD, Cones, FileR} unveil the shape space/relational space variables.  
However the transformations in Sec 2 and 3 of \cite{FileR} for RPM's are purely configuration space manoeuvres, 
whereas passing to Ashtekar variables is itself a canonical transformation.  
Thus RPM's fit better the philosophy of the central importance of the configuration space \cite{B94I, ARel} (as opposed to phase space or of any `polarizations' 
of it that aren't physically configurations).  
More ambitiously, might one be able to extend Halliwell's class operator construction to LQG, so as to be able to, at least formally, write 
down a (perhaps partial) set of complete observables as a subset of the linear constraint complying knots?
[I.e. a tentative search for a fully-fledged Unicorn.]
This goes beyond the currently-known geometrodynamical minisuperspace scope of Halliwell's class operators, but is an interesting direction in which 
to try to extend this scope; this discussion is further motivated by how Schroeren \cite{Schroeren} has recently considered Hartle-type class functions 
for LQG, and is currently investigating Halliwell-type counterparts.  


\noindent {\bf Acknowledgements} I thank Mr Eduardo Serna, Professor Marc Lachi$\grave{\me}$ze-Rey and especially 
Professor Jonathan Halliwell for discussions (and Jonathan also for pointing out various references and complications).  
This work was funded by a grant from the Foundational Questions Institute (FQXi) Fund, 
a donor-advised fund of the Silicon Valley Community Foundation on the basis of proposal FQXi-RFP3-1101 to the FQXi.  
I thank also Theiss Research and the CNRS for administering this grant.  
I thank also Professors Marc Lachi$\grave{\me}$ze-Rey and David Langlois for APC travel money used for part of this project.
I furtherly thank Professor Marc Lachi$\grave{\me}$ze-Rey for help with my career.  
I thank my wife Claire, Amelia, Sophie and Sophie for support, and my friends Anya, Bryony, Hannah, Hettie, Amy, Duke, Zander, Sharon, Lynnette and Becky too.  

\mbox{ }

\noindent{\bf \large Appendix A Various attitudes to environments in Quantum Cosmological models}  

\mbox{ } 

\noindent For the RPM's considered, scale dominates shape so as to model the small inhomogeneous fluctuations of the cosmological arena.
There is then a fork as to whether to model this with a notion of environment. 


\noindent Attitude 1) On the one hand, the 3-particle RPM has hitherto been taken as a whole-universe model.  
This most ideal interpretation is a lot less robust to assuming existence of additional particles whose contributions are then to be traced over 
(as compared with  minisuperspace modelling having little problem with inclusion of an environment since one assumes there that this model sits 
in some kind of neglected environment of small inhomogeneous fluctuations).  
The issue then is justifying the latter parts of \cite{H03, HT} (or, even more so, the elsewise more correct \cite{H09, H11}) in the absense of an environment; this costs us e.g. 
lines (\ref{pipupi}-\ref{whiffle}) in the triangleland example. 
\noindent I note however that in the Halliwell 2009 parallel, configuration space regions that are large enough need no environment.\footnote{Here,  
large means compared to wavelength, which is solution-dependent. 
I also note that spherical geometry cases of this as per regions of the shapes of Fig \ref{Triada}c) are not really any harder to treat than the usual flat space cases.}
%
As such, our next port of call (Paper II) is the RPM counterpart of Halliwell 2009 rather than further discussions of environment or its lack in the Halliwell 
2003 context.  


\noindent Attitude 2) Alternatively, study scale alone and use shape as an environment.   
This suffers from over-simpleness of the original system, but this is alleviated once one considers the shape perturbations about this.  


\noindent Attitude 3) Study a small set of particles (say a triangle of particles) that are taken to dominate over the cloud of other 
particles, which make small and averaged-out contributions.  
These other particles are taken to be negligible in terms of most of their physical effects, but are still permitted to serve as an 
environment for decoherence and accompanying approximate information storage.
Even 1 particle's worth of environment can serve as a nontrivial environment \cite{H99}.    
This offers a second resolution to `losing \cite{HT}'s environment': arguing that it was hitherto negligible in the study but is 
nevertheless available at this stage of the study as an environment.  


\end{document}